\documentclass[12pt]{article}
\usepackage{amsmath}
\usepackage{graphicx}
\usepackage{enumerate}
\usepackage{natbib}
\usepackage{url} 

\newcommand{\blind}{1}

\usepackage{amsthm,amssymb,color}
\usepackage{times}
\usepackage{epstopdf}
\usepackage{bm}
\usepackage{booktabs}
\usepackage[usenames,dvipsnames]{xcolor}
\usepackage{tikz} 
\usepackage[plain,noend]{algorithm2e}
\usetikzlibrary{positioning}
\usepackage[shortlabels]{enumitem}

\newcommand{\SFSred}[1]{#1}
\newcommand{\SFS}[1]{#1}
\newcommand{\SFSnew}[1]{#1}
\newcommand{\SFSnewRed}[1]{#1}

\usepackage{hyperref}
\hypersetup{
    colorlinks,
    linkcolor={red!50!black},
    citecolor={blue!50!black},
    urlcolor={blue!80!black}
}

\usepackage[title]{appendix}
\usepackage{colortbl}

\newtheorem{theorem}{Theorem}
\newtheorem{corollary}{Corollary}[theorem]

\theoremstyle{definition}

\newtheorem{alg}{Algorithm}

\newcommand{\apsi}{a}
\newcommand{\atau}{a} 

\newcommand{\betac}{\beta} 
\newcommand{\betaci}[1]{{\beta}_{#1}} 
\newcommand{\betad}{d} 
\newcommand{\Betadis}[1]{\mathcal{B}\left(#1\right)}
\newcommand{\Betafun}[1]{B (#1)}
\newcommand{\betai}[1]{{\betav}_{#1}} 
\newcommand{\Betapr}[1]{\mathcal{BP}\left(#1\right)}
\newcommand{\GenBetapr}[1]{\mathcal{BP}'\left(#1\right)}
\newcommand{\betar}{\boldsymbol{\betac}} 
\newcommand{\betav}{\betar} 

\newcommand{\bfz}{{\mathbf{0}}}
\newcommand{\bfzmat}{{\mathbf{O}}}
 
 \newcommand{\btildeci}[1]{\tilde{\beta}_{#1}} 

\newcommand{\Chi}{\chi^2}
\newcommand{\Chisqu}[1]{\chi^2 _{#1}}
\newcommand{\Chinon}[2]{\chi^2 _{#1;#2}}
\newcommand{\Chinoncen}{\Lambda}

\newcommand{\cpsi}{c}
\newcommand{\ctau}{c} 

\newcommand{\Diag}[1]{\mbox{\rm Diag}\left(#1\right)}
\newcommand{\dimmat}[2]{(#1\times #2)} 

\newcommand{\e}{\mbox{\rm e}}
\newcommand\equald{\,{\buildrel d \over =}\,}
\newcommand{\error}{\varepsilon} 

\newcommand{\Fd}[1]{\mbox{\rm F}\left(#1\right)}

\newcommand{\Fdnon}[2]{\mbox{\rm F}\left(#1;#2\right)}

\newcommand{\Gammad}[1]{ \mathcal{G}{{\small \left(#1\right)}}}
\newcommand{\Gamfun}[1]{\Gamma (#1)}

\newcommand{\Gammainv}[1]{\mathcal{G}^{-1} \left(#1\right)}

\newcommand{\iid }{\mbox{\rm i.i.d.}}

\newcommand{\indic}[1]{\mathbb{I} \{#1\}}

\newcommand{\kappaBj}[1]{\kappa_{#1}}

\newcommand{\kappaint}{\phi}
\newcommand{\kappaM}{\boldsymbol{\kappaPG}}
\newcommand{\kappaPG}{\kappa}

\newcommand{\kfQc}{\theta}  
\newcommand{\kfQ}{{\mathbf{Q}}}  
\newcommand{\kfwc}{w}

\newcommand{\kfw}{{\mathbf{\kfwc}}}  

\newcommand{\Kmax}{K_{\footnotesize \max}}

\newcommand{\lambdac}{\lambda} 

\newcommand{\minusindex}[1]{ - #1 }
\newcommand{\minusmcmc}[1]{_{\minusindex{#1}}} 
\newcommand{\muh}{\mu^h}

\newcommand{\NegBin}[1]{\mbox{\rm NegBin}\left(#1\right)}

\newcommand{\Normal}[1]{ \mathcal{N}\left(#1\right)}
\newcommand{\Normult}[2]{ \mathcal{N} _{#1}\left(#2\right)}

\newcommand{\phih}{\phi^h}
\newcommand{\Phim}{\boldsymbol{\Phi}}
\newcommand{\phiTG}{\phi}

\newcommand{\piNB}{\pi}
\newcommand{\pkappa}{P}
\newcommand{\pkappam}{\mathbf{\pkappa}}
\newcommand{\Poi}[1]{\mathcal{P}\left(#1\right)}
\newcommand{\Poisson}[1]{\Poi{#1}}
\newcommand{\Probsym}{\mbox{\rm Pr}}
\newcommand{\Prob}[1]{\Probsym (#1)}
\newcommand{\psis}{\psi}
\newcommand{\psitilde}{\tilde{\psi}}

\newcommand{\zm}{\bm z}
\newcommand{\zcond}[1]{\zm  \minusmcmc{#1}}

\newcommand{\Qrcm}{{\mathbf{\kfQ}}}

\newcommand{\rhotr}{\rho}
\newcommand{\rvX}{X} 
\newcommand{\rvXx}{x} 
\newcommand{\rvY}{Y} 
\newcommand{\rvYy}{y} 

\newcommand{\sigmaerr}{\sigma^2}

\newcommand{\Student}[2]{t _{#1} \left(#2\right)}

\newcommand{\thetatr}{\vartheta} 

\newcommand{\Um}{{\mathbf{U}}}

\newcommand{\wt}[1]{\kfw_{#1}}

\newcommand{\Xbeta}{{\mathbf \Xz}}

\newcommand{\Xz}{x}

\newcommand{\Zd}[1]{\mbox{\rm Z}\left(#1\right)}
\newcommand{\Uhyp}[1]{U \left(#1\right)}

\usepackage{csquotes}
\MakeOuterQuote{"}

\begin{document}

\def\spacingset#1{\renewcommand{\baselinestretch}%
{#1}\small\normalsize} \spacingset{1}


\if1\blind
{
  \title{\bf The Dynamic Triple Gamma as a Shrinkage Process for Time-Varying Parameter Models}
  \author{Peter Knaus\hspace{.2cm}\\
    Department of Statistics, Harvard Univertsity\\
    and \\
    Sylvia Frühwirth-Schnatter\\
    Institute for Statistics and Mathematics, WU Vienna}
  \maketitle
} \fi

\if0\blind
{
  \bigskip
  \bigskip
  \bigskip
  \begin{center}
    {\LARGE\bf The Dynamic Triple Gamma as a Shrinkage Process for Time-Varying Parameter Models}
\end{center}
  \medskip
} \fi

\bigskip
\begin{abstract}
Many existing shrinkage approaches for time-varying parameter (TVP) models assume constant innovation variances across time points, inducing sparsity by shrinking these variances toward zero. However, this assumption falls short when states exhibit large jumps or structural changes, as often seen in empirical time series analysis. To address this, we propose the dynamic triple gamma prior -- a stochastic process that induces time-dependent shrinkage by modeling dependence among innovations while retaining a well-known triple gamma marginal distribution. This framework encompasses various special and limiting cases, including the horseshoe shrinkage prior, making it highly flexible. We derive key properties of the dynamic triple gamma that highlight its dynamic shrinkage behavior and develop an efficient Markov chain Monte Carlo algorithm for posterior sampling. The proposed approach is evaluated through sparse covariance modeling and forecasting of the returns of the EURO STOXX 50 index, demonstrating favorable forecasting performance.
\end{abstract}

\noindent%
\vfill

\newpage
\spacingset{1.9} 
\section{Introduction} \label{sec_intro}

It has become evident in recent years that time-series methods must be able to adapt to large, sudden changes in underlying system dynamics. However, allowing too much flexibility can lead to overfitting and poor out-of-sample performance. Therefore, a balance must be struck between local adaptivity and regularization. Time-varying parameter (TVP) models provide a natural framework for exploring this balance, as they are inherently flexible and thus require proper regularization. 

In the Bayesian paradigm, regularization is often achieved through shrinkage priors, which are mostly designed with variable selection in mind. However, regularization in TVP models is inherently a problem of \textit{variance} selection, as it necessitates pulling innovation variances toward zero. Within the TVP framework, \cite{fru-wag:sto} addressed this limitation by re-casting variance selection as a variable selection problem using a non-centered parameterization of the standard TVP model. In this reparameterized state-space model, shrinkage priors can be adapted to handle variance selection with minimal modification. 

\begin{figure}[t!]
\vspace{-1em}
\begin{center}
\includegraphics[width=0.85\textwidth]
{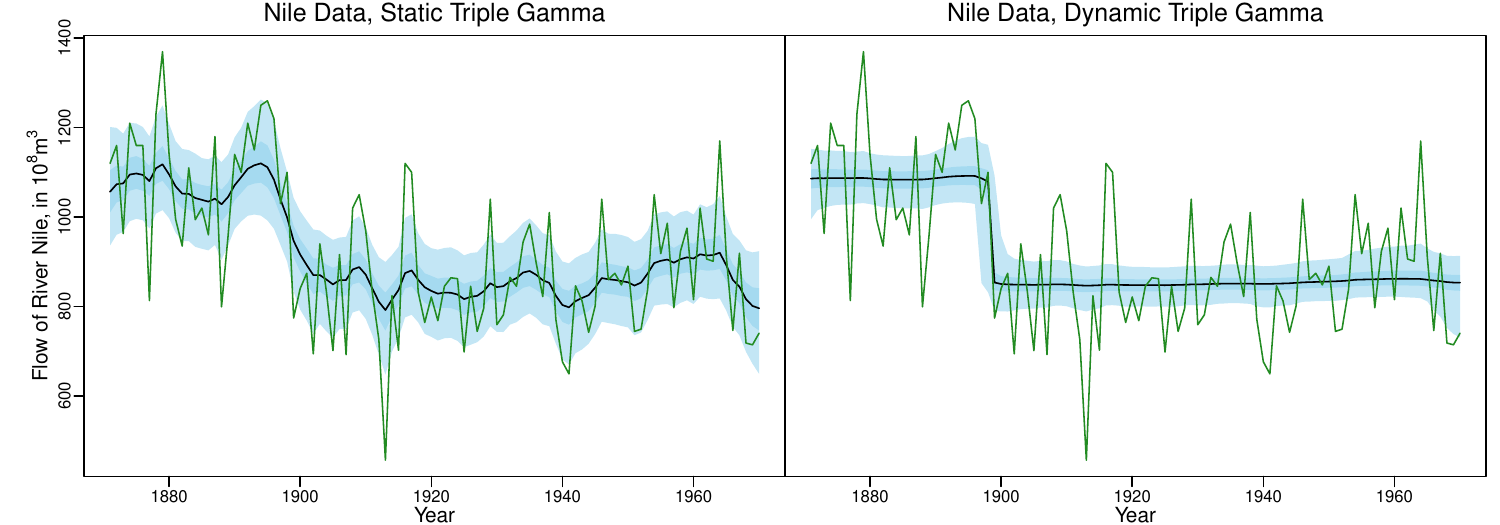}
\caption{\small Posterior estimate of a time-varying mean model fit to Nile data set under static triple gamma and dynamic triple gamma (with $\apsi = \cpsi = 0.5$). Black lines represent posterior medians, while shaded regions represent point-wise 50\% and 95\% credible intervals. The green lines represent the observations.} \label{fig_shortcomings}
\end{center}
\vspace{-1em}
\end{figure}

This approach to shrinkage is inherently \lq\lq static\rq\rq , as the amount of shrinkage applied to each state never changes over time. Therefore, it may simultaneously suffer from overfitting and underfitting. To see this, consider the well-known Nile data set from \cite{kraus1956graphs} displayed in Figure~\ref{fig_shortcomings}.
It consists of 100 measurements of the annual flow of the river Nile at Aswan, with an apparent change-point near 1898, most likely due to a change in rainfall patterns. Such a change-point problem necessitates local adaptivity, as it requires a limited number of large jumps (i.e. 
\SFS{innovations with}
high variance) followed and preceded by periods of 
\SFS{innovations with}
low variance. The left-hand panel of Figure~\ref{fig_shortcomings} plots the posterior of a time-varying mean model\footnote{Details on the exact model specification and priors used 
can be found in Appendix~\ref{sec_local_level}.} using an approach with a constant innovation variance. The aforementioned underfitting is apparent around the change-point, as the shift downwards is quite gradual, where it should be fairly sharp. Overfitting, on the other hand, can be seen to the left and right of the change-point, where it seems a lot of noise is tarnishing the estimate of the mean of the time series. Contrast this with the right-hand panel of Figure~\ref{fig_shortcomings}, which graphs the posterior of a time-varying mean model 
under the proposed dynamic shrinkage method, which nicely picks up the change point while not suffering from the aforementioned overfitting in the less volatile periods.

This limitation of static shrinkage approaches has been widely recognized, leading to a wealth of literature that aims to develop dynamic shrinkage.
An early discussion of dynamic linear models with time-varying variance can be found in \cite{pet-etal:dyn}. Further pioneering work in this direction was undertaken by \cite{koo-kor:for}, who developed a time-varying Bayesian model averaging scheme and \cite{nak-wes:bay_ana}, who used a latent threshold approach to induce time-varying sparsity. Many other approaches are based on the idea of generalizing a shrinkage prior designed for variable selection to handle dynamic shrinkage. Due to its attractive theoretical properties and interpretability, the spike-and-slab approach \citep{geo-mcc:var} is a popular candidate for extension, with work in this direction done by \cite{hub-etal:sho}, \cite{uri-hed:dyn}, and \cite{roc-mca:dyn}, among others. 
\SFSnew{Another }strand of the literature is defined by a continuous shrinkage prior serving as the basis for generalization into a shrinkage process prior, with examples including work by \cite{kal-gri:tim}, \cite{kow-etal:dyn}, \cite{irie2019bayesian} and \cite{hub-pfa:dyn}. 
\SFS{Recent machine learning approaches include}
 the mixture approach by \cite{hau-etal:fas} or the tree-based contribution of \cite{hau-etal:tree}.  

This paper generalizes the triple gamma shrinkage prior of \cite{cad-etal:tri} to a stochastic process that also captures dependence in the amount of shrinkage across time. The triple gamma’s spike captures the belief that most innovations are near zero, while its heavy tails still allow for large jumps. Dependence enables
volatility information 
to be shared across time. 
Since the triple gamma encompasses many popular shrinkage priors as special or limiting cases, this defines a class of stochastic shrinkage processes with diverse shrinkage priors as marginals. 
%
\SFSnew{The most prominent example is an innovative shrinkage process prior that captures dependence 
in the amount of shrinkage across time 
and features the famous horseshoe prior \citep{car-etal:hor} as marginal distribution;}
a combination that, to the best of our knowledge, is novel within the existing literature. Despite these features, the full conditional distributions of the posterior remain mostly in closed form, simplifying estimation.

The rest of the paper is structured as follows: Section~\ref{sec_shr} lays groundwork by providing a brief overview how  
global-local (GL) shrinkage priors \SFSnew{can be applied to shrinking innovation variances} in TVP models \SFSnew{toward zero}. Section~\ref{sec_ds} introduces the key contribution, 
\SFSnew{namely a TVP model with time-varying innovation variances following the innovative shrinkage process defined by a dynamic triple gamma.}
Section~\ref{sec_dtg} discusses properties of the dynamic triple gamma prior in more detail, while Section~\ref{compu} develops an efficient Markov chain Monte Carlo (MCMC) sampler to estimate TVP models under the dynamic triple gamma. Section~\ref{appl} showcases the dynamic triple gamma by fitting a Cholesky stochastic volatility (SV) model to the returns of the components of the EURO STOXX 50 index and demonstrating the out-of-sample forecasting performance via log predictive density scores (LPDS). Finally, Section~\ref{conc} concludes.

\section{Shrinking \SFSnew{Innovation} Variances in TVP Models} \label{sec_shr}

To set notation for the following discussion, we introduce the standard TVP model:
\begin{eqnarray}
 & & y_{t} =   \Xbeta_{t} \betai{t} +  \error_{t} , \qquad
\error_{t} \sim \Normal{0,\sigmaerr_t}, \label{lgssm:se} \\
&&  \betai{t} = \betai{t-1} + \wt{t}, \qquad   \wt{t}  \sim \Normult{\betad}{\bfz,\Qrcm }, \label{ms:ARPmgen}
\end{eqnarray}
where the $\dimmat{1}{\betad}$-dimensional \SFSnew{row} vector $\Xbeta_t  
$ contains the regressors and $y_t$ is the regressand, for $t = 1, \dots T$. 
\SFSnew{The innovations' covariance matrix}
$\Qrcm  =\Diag{\kfQc_1, \ldots, \kfQc_{\betad}}$ 
\SFSnew{is  diagonal,  
meaning that} 
each coefficient follows a random walk with innovation 
variance $\kfQc_j$, i.e. 
\SFSnew{for $j=1,\ldots, \betad$,}
$\betaci{jt} = \betaci{j,t-1} + \kfwc_{jt}$, where $\kfwc_{jt} \sim \Normal{0 ,\kfQc_j}$.
The system is initialized by  assuming that the initial value  $ \betai{0}$ is randomly drawn from the following normal distribution: $ \betai{0} \sim  \Normult{\betad}{\betav,\Qrcm }$, where $\betav = (\beta_1, \dots, \beta_d)'$. The error variance 
can either be assumed to be homoscedastic or follow 
\SFSnew{the SV model of \cite{jac-etal:bay_ana} where  $\sigmaerr_t = \e ^ {h_t}$ and $h_t$ follows an AR(1) process.}


These models are inherently flexible, as the regression coefficients can change at each time point $t \in \{1, \dots, T\}$. One approach to regularizing such models is to shrink the innovation variances $\kfQc_1, \dots, \kfQc_d$ toward zero, as smaller variances indicate that innovations are, on average, closer to zero. In the extreme case, where $\kfQc_j = 0$, we have $\beta_{jt} = \beta_{j,t-1} = \beta_j$ with probability 1. 
\SFSnew{While}  GL shrinkage priors are \SFSnew{primarily}    designed for variable selection, 
\SFS{they can also be used to} address variance selection in 
\SFSnew{a TVP model. To this end, }
\cite{fru-wag:sto}
\SFSnew{introduced following}
 useful reparameterization of the standard TVP model:
\begin{eqnarray}
 & & y_{t} = \Xbeta_{t}\betav + \Xbeta_{t} \Diag{\sqrt\kfQc_1, \dots, \sqrt\kfQc_d} +  \Xbeta_{t} \Tilde{\betai{t}} +  \error_{t} , \qquad
\error_{t} \sim \Normal{0,\sigmaerr_t}, \label{lgssm:se_noncen} \\
&& 
\SFS{\btildeci{jt} = \btildeci{j,t-1} + \tilde{\kfwc}_{jt}, \quad \tilde{\kfwc}_{jt} \sim \Normal{0 ,1}, \qquad j=1,\ldots, \betad, }
\label{ms:ARPmgen_noncen}
\end{eqnarray}
where
\SFS{$\btildeci{j0} \sim \Normal{0 ,1}$.}
The two parameterizations are linked through the simple transformation $\beta_{jt} = \beta_{j} + \sqrt\kfQc_j \Tilde{\beta}_{jt}$, for all $t = 1, \dots, T$. To see the benefit of this, consider placing a generic GL shrinkage prior on $\sqrt \kfQc_j$,
where 
\SFS{$\sqrt\kfQc_j | \lambda_j, \tau \sim \Normal{0, \tau\lambda_j}$},
$\tau$ is the global parameter pulling all coefficients toward zero and 
$\lambda_j \sim \SFS{p(\lambda_j)}$ 
is the local parameter that allows individual coefficients to be different from zero.
When such a prior is placed on $\sqrt\kfQc_j$, it implies the following GL variance shrinkage prior on $\kfQc_j$ 
\SFS{based on the gamma distribution}:
\begin{equation*}    
\kfQc_j | \lambda_j, \tau \sim \Gammad{\frac 1 2, \frac 1 {2\tau\lambda_j}}, 
\quad \lambda_j \sim \SFS{p(\lambda_j)}.
\end{equation*}
This duality is what makes this reparameterization so useful -- any GL prior originally desigend for variable selection can be placed on $\sqrt \kfQc_j$ and will automatically imply a shrinkage prior for 
\SFS{the innovation variances  $\kfQc_j$ in a TVP model.}
This is the context in which the triple gamma prior was introduced by \cite{cad-etal:tri}. 
%


In this paper, we extend the standard TVP model by allowing for dynamic innovation variances. Specifically, each innovation variance \( \kfQc_j \) is multiplied by a local, time-varying component \( \psis_{jt} \), resulting in \(\wt{t} \sim \Normult{\betad}{\bfz, \Qrcm_t}\) with \(\Qrcm_t = \Diag{\kfQc_1 \psis_{1t}, \ldots, \kfQc_{\betad} \psis_{\betad t}}\). It will be shown that the triple gamma prior is also a natural choice as a shrinkage prior for \( \psis_{jt} \) in achieving dynamic shrinkage.

\subsection{\SFSnew{A brief Review} 
of the Triple Gamma Shrinkage Prior}
\label{sec_properties}


\SFSnew{The triple gamma shrinkage prior $TG\left(a, c, \kappa\right)$ is a distribution 
with three hyperparameters, where $a$ controls the behavior at 0, $c$ controls the tail behavior and
$\kappa$ controls global shrinkage.} 
The eponymous representation of the triple gamma is a compound distribution consisting of three gamma random variables, meaning that $ X | a, c, \kappa \sim TG\left(a, c, \kappa\right)$ is equivalent to
\begin{equation*}
 X | Y, \kappa \sim \Gammad{\frac{1}{2}, \frac{\kappa}{2Y}}, \quad Y|Z, a \sim \Gammad{a, \frac{a Z}{2}}, \quad Z|c \sim \Gammad{c, c}.
\end{equation*}
It can also be represented in a form closer to the generic GL representation, using the connection between the gamma and the F distribution:
\begin{equation*} \label{TG_GL}
 X | W, \kappa \sim \Gammad{\frac{1}{2}, \frac{\kappa}{4W}}, \quad W | a, c \sim \Fd{2a, 2c}.
\end{equation*}
It's variable selection "twin" is equivalent to the normal-gamma-gamma \SFSnew{(NGG)} 
prior of \cite{gri-bro:hie}:
\begin{equation*} \label{NGG}
 \sqrt X | Y, \kappa \sim \Normal{0, \frac 2 \kappa Y}, \quad Y|Z, a \sim \Gammad{a, \frac{a Z}{2}}, \quad Z|c \sim \Gammad{c, c},
\end{equation*}
or, in GL form:
\begin{equation} \label{NGG_GL}
 \sqrt X | W, \kappa \sim \Normal{0,  \frac 2 \kappa W}, \quad W | a, c \sim \Fd{2a, 2c}.
\end{equation}
This prior is 
appealing for two reasons. First, the triple gamma prior either is equivalent to or encompasses many popular GL priors as special or limiting cases. 
As such, it can be considered a fairly generic choice of shrinkage prior, as many popular choices result 
\SFSnew{for specific values of $a$ and $c$, the most prominent example being the horseshoe prior for $a=c=0.5$. This is evident from \eqref{NGG_GL}, since $W \sim \Fd{1,1}$ implies 
that $\sqrt{W} \sim t_{1}$ follows a Cauchy distribution.}
Second, it is mathematically 
well understood, in particular with respect to it's behavior around the origin and in the tails. This results from the marginal density being available in closed form:
\begin{equation} \label{tg_def}
    f(x; a, c, \kappa) = \frac{\Gamfun{c+{\frac{1}{2}}}}{\sqrt{2\pi \phi x} \Betafun{a,c}}
		\Uhyp{c + { \frac{1}{2}}, \frac{3}{2} - a, \frac{x}{2 \phi} },
\end{equation}
where $\phi = \frac{2c}{\kappa a}$ and $\Uhyp{a,b,z}$ is the confluent hypergeometric function of the second kind:
\begin{align*}
\Uhyp{a,b,z} = \frac{1}{\Gamma(a)}\int_0^\infty e^{-zt} t^{a-1}(1+t)^{b-a-1} dt.
\end{align*}

\begin{figure}[t!]
\vspace{-1em}
\begin{center}
\includegraphics[width=0.85\textwidth]
{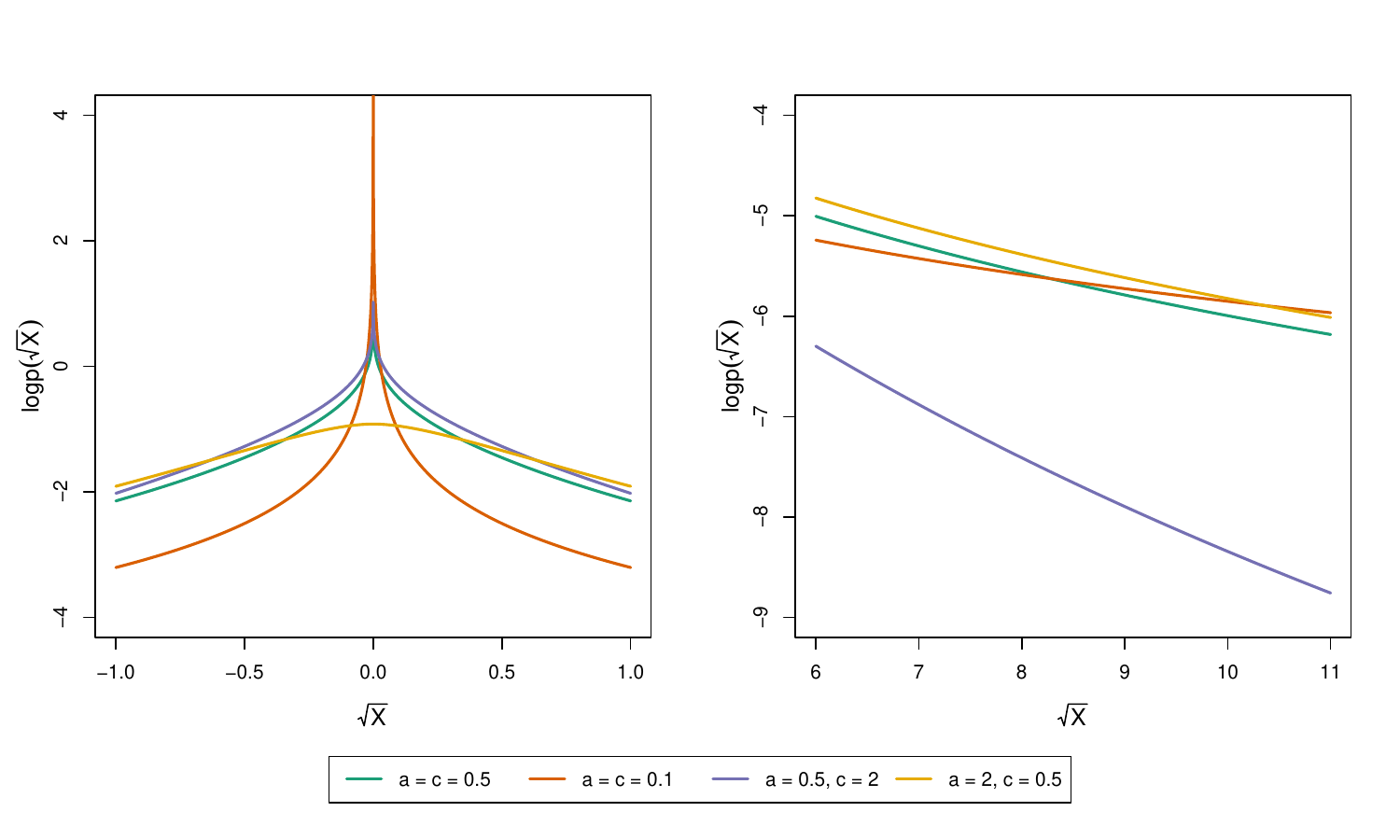}
\caption{\small  Log 
\SFSnew{probability density function} 
of the triple gamma prior for different values of $a$ and $c$. 
\SFS{Smaller} 
values of $a$ lead to a more pronounced pole around $0$; 
smaller values of $c$ lead to heavier tails.} \label{fig_a_c}
\end{center}
\vspace{-1em}
\end{figure}

\noindent The properties of the confluent hypergeometric function \SFS{can be} used to characterize the pole and tail behavior \SFS{\citep[Theorem~2]{cad-etal:tri}.
The} triple gamma prior has an infinite pole around $0$ when $a \leq 0.5$ and 
this pole becomes more extreme the smaller $a$ becomes. 
\SFS{Furthermore, }
the prior has polynomial tails, with smaller values of $c$ equating to heavier tails. This is further reinforced by the fact that the prior moments $\mathbb E(X^k|a, c, \kappa)$ only exist up to $k < 2c$. Interestingly, the horseshoe in particular results as the special case when $a = c = 0.5$. As such, it has the least mass around the origin of all triple gamma specifications with an infinite pole at $0$, while being the most heavy-tailed the triple gamma can be while still still having a finite first moment. If a reader is so inclined, they can therefore think of the triple gamma as a more flexible horseshoe prior that allows for more granular control over the behavior around the origin and in the tails. A visual demonstration of this can be found in Figure~\ref{fig_a_c}, which shows  the log density of the variable selection twin of the triple gamma prior, defined on the entire real line. The figure illustrates that specifications with the same $a$ behave very similarly around the origin, while those with the same $c$ exhibit similar behavior in the tails. Finally, 
\SFS{$ \tau=2/\kappa$}
acts as the global shrinkage component, with larger values of $\kappa$ corresponding to more overall shrinkage.

\subsection{\SFSnew{Exchangeable Triple Gamma Innovations}} \label{sec_exTG}

\SFSnew{To introduce dynamic shrinkage through the triple gamma prior, we begin  with an} exchangeable prior 
on the innovations and then extend it 
\SFSnew{in Section~\ref{sec_ds}}
to the case that allows for dependence across time. 
\SFSnew{As mentioned above, 
we extend the} TVP model defined in \eqref{lgssm:se} - \eqref{ms:ARPmgen}, with a triple gamma prior
$\kfQc_j \sim TG ( a^\xi, c^\xi, \kappaBj{B}^2)$ on the innovation variance,
 by multiplying  $\kfQc_j$ with a local component $\psis_{jt}$.  
\SFSnew{If $\psis_{jt}<1$, then global shrinkage induced for coefficient $j$ by $\kfQc_j$ will be enforced, while $\psis_{jt}>>1$ will apriori allow for large local innovations $\kfwc_{jt} $.}
While \SFSnew{other} 
choices for the distribution of $\psis_{jt}$ are feasible, \SFSnew{the easiest is to} use following 
exchangeable prior:
\begin{equation} \label{iid_f}
     \kfwc_{jt} \SFS{|\kfQc_j,\psis_{jt}} \sim \Normal{0, \kfQc_j\psis_{jt}}, 
     \quad \psis_{jt} \SFS{|\apsi_j ,  \cpsi_j }
\,\, \mathop{\sim}\limits^{\mathrm{iid}} \,\,
     \Fd{2\apsi_j, 2\cpsi_j}.
\end{equation}
This choice leads to $\kfwc_{jt}^2|
\SFSnew{\apsi_j, \cpsi_j, \kfQc_j}$ being distributed as \iid\ $TG(\apsi_j, \cpsi_j, 2/\kfQc_j)$, which follows immediately from representation \eqref{NGG_GL}. A similar exchangeable approach to dynamic shrinkage can be seen in
\SFS{\cite{hub-etal:ind}},
where they place exchangeable horseshoe priors on the innovations.

The structure of this setup can be viewed as a doubly hierarchical GL prior on the innovations. The $\kfQc_j$'s serve as the global component within each state, shrinking all innovations $\kfwc_{jt}$ within state $j$ (for $t = 1, \dots, T$), while the $\psis_{jt}$'s act as the local components. Further, the $\kfQc_j$'s themselves are hierarchical, with following representation
\SFS{involving local components $\xi_j$:}
\begin{equation*}
    \kfQc_j | \xi_j, \kappaBj{B}^2
    \sim \Gammad{\frac 1 2, \frac{\kappaBj{B}^2}{4\xi_j}}, \quad \xi_j | a^\xi, c^\xi 
    \,\, \mathop{\sim}\limits^{\mathrm{iid}} \,\,
    \Fd{2a^\xi, 2c^\xi}.
\end{equation*}
Thus, $\kfQc_j$ controls the variability within state $j$, while $\kappaBj{B}^2$ regulates the range of $\kfQc_j$ across states.

\section{The Dynamic Triple Gamma Prior} \label{sec_ds}

While the exchangeable triple gamma innovations 
\SFSnew{introduced in Section~\ref{sec_exTG}}
are interesting in and of themselves, they do not yet allow information on the 
\SFSnew{amount of shrinkage}
to be shared across time points. However, the assumption that high volatility in one period is correlated with higher volatility in the next seems fairly reasonable. 
This requires the local shrinkage factors \( \psis_{jt} \) to follow a stochastic shrinkage process prior, similar to \cite{kal-gri:tim} and \cite{kow-etal:dyn}. At the same time, we aim to preserve the triple gamma prior as the marginal distribution of this process, owing to its well-understood mathematical properties, briefly reviewed in Section~\ref{sec_properties}. This motivates the introduction of the dynamic triple gamma, a novel stochastic shrinkage process prior.

We first begin by stating an alternate, novel representation of the triple gamma prior. We use properties of 
the 
\SFS{$\Fd{2 \apsi_j, 2 \cpsi_j}$-distribution }
to represent $\psis_{jt}$ 
in \eqref{iid_f} as\footnote{To see this, note that the F-distribution   has a representation as the ratio of two independent gamma  distributions, 
$\Fd{2 \apsi_j, 2 \cpsi_j} \equald \frac{\Gammad{\apsi_j,1}/\apsi_j}{\Gammad{\cpsi_j,1}/\cpsi_j} \equald \frac{\Gammad{\apsi_j,1}\cpsi_j/\apsi_j}{\Gammad{\cpsi_j,1}}$,
which can be written as
$\Fd{2 \apsi_j, 2 \cpsi_j} \equald \Gammainv{\cpsi_j,\Gammad{\apsi_j,\apsi_j/\cpsi_j}}$.}
\begin{eqnarray} \label{rephor}
 \SFS{\kfwc_{jt} \SFS{|\kfQc_j,\psis_{jt}} \sim \Normal{0, \kfQc_j\psis_{jt}},} 
     \,\, \psis_{jt}|  \lambdac_{jt}, \cpsi_j \sim \Gammainv{\cpsi_j , \lambdac_{jt} } , \quad \lambdac_{jt} | \apsi_j ,  \cpsi_j   \mathop{\sim}\limits^{\mathrm{iid}}
     \Gammad{\apsi_j, \frac{\apsi_j}{\cpsi_j}}  .
\end{eqnarray} 
Interestingly, this alternative representation also leads to a novel sampling scheme for the posterior 
under triple gamma  priors, as can be seen in the MCMC sampler introduced in Section~\ref{compu}.

The most innovative step in constructing the dynamic triple gamma prior is replacing the exchangeable prior distribution for $ \lambdac_{jt} $ in \eqref{rephor} with a stochastic process that introduces autocorrelation, while preserving the marginal distribution $ \lambdac_{jt} \mid \apsi_j, \cpsi_j \sim \Gammad{\apsi_j, \apsi_j / \cpsi_j} $ as in the exchangeable case. This ensures the marginal distribution of \( \kfwc_{jt} \) remains unchanged. To define such a process, we place the following latent autoregressive gamma (ARG) process on \( \lambdac_{jt} \):  
\begin{eqnarray*}
  & \displaystyle \lambdac_{jt} |\kappaPG_{jt}, \apsi_j, \cpsi_j, \rhotr _j \sim \Gammad{\apsi_j  + \kappaPG_{jt} , \frac{\apsi_j}{\cpsi_j}\frac{1}{1- \rhotr_j }}, &\\
  &  \displaystyle \kappaPG_{jt}|\lambdac_{j,t-1}, \apsi_j, \cpsi_j ,\rhotr _j  \sim \Poisson{\frac{\apsi_j}{\cpsi_j} \frac{\rhotr _j}{1- \rhotr _j} \lambdac_{j,t-1}}, &
\end{eqnarray*}
which was introduced by \cite{gou-jas:aut} as an AR-type process for non-negative time series observations.
\SFSnew{This definition induces the marginal distribution $\lambdac_{jt}|\apsi_j, \cpsi_j \sim \Gammad{\apsi_j, {\apsi_j}/{\cpsi_j}}$. At the same time,} it leads to a process that is autocorrelated due to the dependence of $\lambdac_{jt}$ on $\lambdac_{j,t-1}$ through the contemporaneous value of $\kappaPG_{jt}$, with the strength of the autocorrelation governed by the parameter $\rhotr_j$. More precisely, this construction causes the expectation to be linearly dependent on the previous value, i.e. $\mathbb E[\lambdac_{jt}|\lambdac_{j,t-1}] = \rhotr_j \lambdac_{j,t-1}$. Note that, as $\lambdac_{jt}$ has support on $\mathbb R_{+}$, this only allows for values of $\rhotr_j$ in $(0, 1)$.  To clarify the hierarchical structure of the dynamic triple gamma prior, Figure~\ref{tg_graph} provides a visualization, and the complete construction is presented below: 
\begin{eqnarray}
 && \kfwc_{jt}|\kfQc_j,\psis_{jt} \sim \Normal{0, \kfQc_j\psis_{jt}}, \label{innov} \\
 && \psis_{jt}|  \lambdac_{jt} , \cpsi_j \sim \Gammainv{\cpsi_j , \lambdac_{jt} } , \label{psi}\\
 && \lambdac_{jt} |\kappaPG_{jt}, \apsi_j, \cpsi_j, \rhotr _j \sim \Gammad{\apsi_j  + \kappaPG_{jt} , \frac{\apsi_j}{\cpsi_j}\frac{1}{1- \rhotr_j }}, \label{lambda} \\
  && \kappaPG_{jt}|\lambdac_{j,t-1}, \apsi_j, \cpsi_j ,\rhotr _j  \sim \Poisson{\frac{\apsi_j}{\cpsi_j} \frac{\rhotr _j}{1- \rhotr _j} \lambdac_{j,t-1}} \label{kappa}.
\end{eqnarray}
Analogous to the exchangeable case, this construction forms a doubly hierarchical GL prior. $\kfQc_j$ still acts as the global component within each state $j$, pulling all 
innovation variances 
toward zero, while the hierarchical 
\SFS{prior}
placed on $\kfQc_j$ remains unchanged, inducing a joint prior over all $\kfQc_j$'s. The key distinction is that the local components, the $\psis_{jt}$'s, now explicitly model dependence across time. Despite this, the marginal distribution of 
$\kfwc_{jt}^2$ remains triple gamma.

\begin{figure}[t!]
\vspace{-1em}
\centering
\scalebox{0.75} {
\begin{tikzpicture}[node distance = {5mm}, 
                    main/.style = {draw, circle},  
                    minimum size = 1.4cm] 
                    
\node[main] (1) {$\kfwc_{j,t-1}$};
\node[main, right = of 1] (2) {$\kfwc_{jt}$};
\node[main, right = of 2] (3) {$\kfwc_{j,t+1}$};

\node[main] (4) [below = of 1] {$\psis_{j,t-1}$};
\node[main] (5) [below = of 2] {$\psis_{jt}$};
\node[main] (6) [below = of 3] {$\psis_{j,t+1}$};

\node[main] (7) [below = of 4] {$\lambdac_{j,t-1}$};
\node[main] (8) [below = of 5] {$\lambdac_{jt}$};
\node[main] (9) [below = of 6] {$\lambdac_{j,t+1}$};

\node[main] (10) [below = of 7] {$\kappaPG_{j,t-1}$};
\node[main] (11) [below = of 8] {$\kappaPG_{jt}$};
\node[main] (12) [below = of 9] {$\kappaPG_{j,t+1}$};

\node[main, draw = none] (13) [left = of 7] {};
\node[main, draw = none] (14) [right = of 12] {};

\draw[->] (4) -- (1);
\draw[->] (5) -- (2);
\draw[->] (6) -- (3);

\draw[->] (7) -- (4);
\draw[->] (8) -- (5);
\draw[->] (9) -- (6);

\draw[->] (10) -- (7);
\draw[->] (7) -- (11);
\draw[->] (11) -- (8);
\draw[->] (8) -- (12);
\draw[->] (12) -- (9);

\draw[->] (13) -- (10);
\draw[->] (9) -- (14);

\end{tikzpicture} 
}
\caption{\small Interdependencies of the latent variables of the DTG prior represented as a directed graph.}
\label{tg_graph}
\vspace{-1em}
\end{figure}
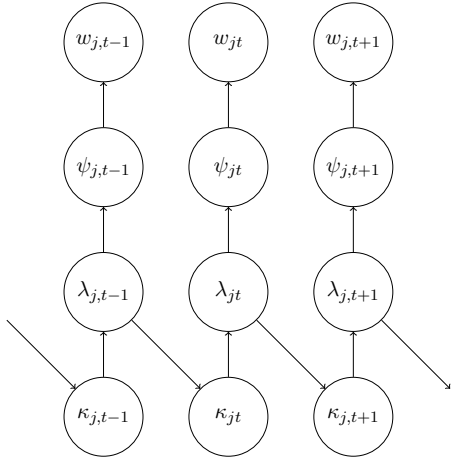

\subsection{\SFS{Hyperprior choices}} 
\label{prior_choices}

From a Bayesian perspective, hyperpriors must be specified for the 
\SFSnew{model parameters} 
$\kfQc_1, \dots, \kfQc_d$, $\betaci{1}, \dots, \betaci{d}$, 
and, crucially, $\rhotr_1, \dots, \rhotr_d$. Since the dynamic triple gamma reduces to the static case when $\psis_{jt} = 1$, we follow \cite{cad-etal:tri} for the parameters shared with  \SFSnew{this} 
model, namely $\kfQc_1, \dots, \kfQc_d$, $\betaci{1}, \dots, \betaci{d}$ and \SFSnew{the error variance $\sigma^2_t$}. 
A key element of this approach is
\SFSnew{employing 
triple gamma priors $\betaci{j}^2 \sim TG(a^\tau, c^\tau, \lambda_B^2)$ and  $\kfQc_j \sim TG(a^\xi, c^\xi, \kappaBj{B}^2)$}
for $j = 1, \dots, d$, allowing the dynamic triple gamma to differentiate between covariates with time-varying, constant, or no effect. Specifically, when $\kfQc_j = 0$, 
\SFSnew{$\betaci{jt} = \betaci{j,t-1}=\beta_j$} with probability 1, reducing the effect to a constant value. 
\SFSnew{The triple gamma prior, equivalent to an NGG prior on \( \betaci{j} \), also pulls static coefficients toward 0.}
If, additionally, $\SFSnewRed{\beta_j = 0}$, the covariate's effect is \SFSnew{constantly zero}, effectively excluding it from the model. \SFSnew{For additional details on these priors   see  Appendix~\ref{sec_prior_details}.}

A prior must also be specified for the 
$\rhotr_j$'s, \SFSnew{model parameters} 
unique to the dynamic case. Since 
\cite{gou-jas:aut} demonstrate that the ARG process degenerates as $\rhotr_j$ approaches 1, we adopt a generalized beta prior of the first kind (GB1) for $\rhotr_j$ to mitigate this issue. Its density function is:
\begin{equation} \label{rho_prior}
f(\rhotr_j ; a_\rho, b_\rho, \alpha_\rho, \beta_\rho) = \frac{|a_\rho| \rhotr_j^{a_\rho \alpha_\rho - 1} \left(1 - (\rhotr_j / b_\rho)^{a_\rho}\right)^{\beta_\rho - 1}}{b_\rho^{a_\rho \alpha_\rho} B(\alpha_\rho, \beta_\rho)},
\end{equation}
valid for $0 < \rhotr_j < b_\rho$, where $a_\rho \in \mathbb{R}$, $b_\rho > 0$, $\alpha_\rho > 0$, and $\beta_\rho > 0$. This prior generalizes the beta distribution in two ways: it allows the support to range from 0 to $b_\rho$, while the parameter $a_\rho$ controls the power applied to the argument. For bounding the prior away from 1, the first property is especially useful, as it restricts the prior's support. For instance, in the application presented in Section~\ref{appl}, we set $b_\rho = 0.95$, limiting the range to $(0, 0.95)$.

\subsection{Relationship of the dynamic triple gamma to other approaches}   

The dynamic triple gamma shares a particularly close connection to two approaches previously proposed in the literature. First, the normal-gamma autoregressive (NGAR) process of \cite{kal-gri:tim} uses essentially the same recursive process as defined by \eqref{lambda}-\eqref{kappa}, albeit with a different parameterization. Through this, they define a 
\SFS{strictly stationary}
process directly on the $\betaci{jt}$'s, such that the resulting marginal distribution is a normal-gamma shrinkage prior \citep{gri-bro:inf}. 
\SFSnew{In our opinion, it seems more natural to induce dynamic shrinkage on the growth process $\kfwc_{jt}=\betai{jt}- \betai{j,t-1}$, as we do in the present paper, as this allows more flexibility in accommodating small and large jumps in $\betai{jt}$ at the same time.}

Second, there exists a (slightly less obvious) connection to the dynamic shrinkage process proposed by \cite{kow-etal:dyn}, where dependence is induced in the innovations through an AR(1) process on the log of the innovation variance:
 \begin{align} \label{TGZ}
\kfwc_{jt}|  h_{jt} \sim \Normal{0,\e ^{h_{jt}}}, \quad  h_{jt} = \phih_j  (h_{j,t-1} - \muh_j) + \eta_{jt},  \quad \eta_{jt} \sim \Zd{\apsi_j,\cpsi_j,\muh_j,1},
\end{align}
where the $\eta_{jt}$ arise from a Z-distribution\footnote{For details on the \SFS{various} non-standard distributions used in this section see Appendix~\ref{nonstand_dist}.}
and $\phi_j^h$ acts as an autocorrelation parameter.
To see how the proposed approach and that of \cite{kow-etal:dyn} are related, note that the triple gamma prior can
also be represented as 
\SFS{\cite[Table~1]{cad-etal:tri}}:
\begin{eqnarray}
&& \kfwc_{jt}|\psitilde_{jt},\apsi_j,\cpsi_j,\kfQc_j  \sim \Normal{0, \phiTG_{j} \psitilde_{jt}}, \quad  \psitilde_{jt}|\apsi_j,\cpsi_j \sim \Betapr{\apsi_j,\cpsi_j},  \label{TgBp}
\end{eqnarray}
where $\Betapr{\apsi_j,\cpsi_j}$ is the beta-prime distribution, $\psitilde_{jt} =  \frac{\apsi_j} {\cpsi_j}\psis_{jt}$, and  $\phiTG_{j} = \kfQc_j \frac{ \cpsi_j}{\apsi_j}$. Based on this, it is possible to verify the following new representation of the triple gamma:
 \begin{align} 
\kfwc_{jt}|  h_{jt} \sim \Normal{0,\e ^{h_{jt}}}, \quad  h_{jt}| \apsi_j,\cpsi_j,\kfQc_j \sim \Zd{\apsi_j,\cpsi_j,\log \phiTG_{j},1}.
\end{align}
Specifically, from representation (\ref{TgBp}) we find that $h_{jt}= 
\SFSnew{\log (\phiTG_{j}\psitilde_{jt}) = }
\log \phiTG_{j} + \log \psitilde_{jt}$, where
$\log \psitilde_{jt} \sim \Zd{\apsi_j,\cpsi_j,0,1}$
and therefore $h_{jt} | \apsi_j,\cpsi_j,\kfQc_j \sim \Zd{\apsi_j,\cpsi_j,\log \phiTG_{j},1}$.\footnote{Note that $ X \sim \Betapr{\atau,\ctau}$ implies $ Y= \log X \sim \Zd{\atau,\ctau,0,1}$ and 
therefore $Y + \mu  \sim \Zd{\atau,\ctau,\mu,1} $.}

Essentially, the \textit{conditional} distribution of $\kfwc_{jt}^2| \apsi_j,\cpsi_j,h_{j,t-1},\mu_j^h, \phi_j^h$ proposed by \cite{kow-etal:dyn} is equivalent to the triple gamma prior. This result is already interesting on its own, as it allows the properties of the triple gamma 
to shed light on the characteristics of the dynamic shrinkage process. It further implies that the two approaches coincide when the respective autocorrelation parameters ($\rhotr_j$ and $\phi_j^h$) are equal to $0$, as they both result in squared innovations that are exchangeable triple gamma, with the same parameters 
\SFS{$\apsi_j$ and $\cpsi_j$, while $\mu_j^h =\log \phi_j$}. 
The key difference is that the marginal form of the proposed approach is again well-known, i.e. triple gamma. This allows the shrinkage characteristics of the entire process to be better understood, while sacrificing some tractability of the conditional distribution. There are also computational advantages, as the MCMC algorithm proposed in Section~\ref{compu} does not need to rely on mixture approximations
\SFS{as in \cite{kow-etal:dyn}}, 
which tend to be expensive to evaluate.

\section{Properties of the Dynamic Triple Gamma} \label{sec_dtg}

To gain further understanding of the dynamic triple gamma
\SFSnew{(DTG)}, one can rely on alternative representations of the  process, which can be achieved by marginalizing out sets of latent variables, such as $\lambdac_{j0}, \ldots, \lambdac_{jT}$ or $\psis_{j1}, \ldots, \psis_{jT}$. For the first such representation,  
\SFS{note that marginalizing 
(\ref{rephor})}
over $\{\psis_{jt}\}$ leads to the following representation
(see also \cite{cad-etal:tri}):
 \begin{eqnarray} \label{reptstud}
&&	\kfwc_{jt}|\lambdac_{jt},\cpsi_j , \SFSnew{\kfQc_j} \sim \Student{2\cpsi_j}{0,\SFSnewRed{\frac{\kfQc_j}{c_j}} 
\lambdac_{jt}}.
 \end{eqnarray}
 Therefore, the innovations conditional on $\lambdac_{jt}$ follow a Student-$t$ distribution with  $2\cpsi_j$ degrees of freedom. This representation also has direct implications for the choice of $\cpsi_j$, as \SFSnewRed{the innovations} $\kfwc_{jt}$ 
 have no \SFSnew{finite} moments whenever $\cpsi_j \leq 0.5$. If $\cpsi_j > 0.5$,  
 \SFSnewRed{the innovations} have expectation zero.  
\SFSnewRed{When $\cpsi_j > 1$, then the conditional variance of 
$\kfwc_{jt}|\cpsi_j, \kfQc_j$
is finite.} From this it can be inferred that a smaller $\cpsi_j$ will lead to a process which allows for larger changes in the states a priori, whereas a larger $\cpsi_j$ will cause the changes in the states to be more regularized. This interpretation is in line with the effect of 
$c$  in the static triple gamma discussed in Section~\ref{sec_properties}.


 Further insights into the behavior of the 
DTG prior 
 can be gleaned by examining the transition density $p(\psis_{jt} | \psis_{j,t-1}, \apsi_j, \cpsi_j, \rhotr_j)$ derived under the assumption that $\lambdac_{j,t-1}$ comes from the stationary distribution of the ARG process. This density is available in closed, albeit not well-known, form in Theorem~\ref{theo_cond}. It allows for impulse response analysis, showcasing how the 
 \SFS{process}  
 responds to a shock 
 while in the steady state. 
 The proof can be found in Appendix~\ref{proofs}.

\begin{theorem}\label{theo_cond}
For the dynamic triple gamma prior defined in  \eqref{innov}-\eqref{kappa}, with $\apsi_j > 0$,  $\cpsi_j > 0$, $0 < \rhotr_j < 1$, and $\lambdac_{j,t-1} \sim \Gammad{\apsi_j, \frac{\apsi_j}{\cpsi_j}} $, the following holds:
\begin{equation}
\label{psi_cond}
\begin{aligned} 
p(\psis_{jt} &| \psis_{j,t-1}, \apsi_j, \cpsi_j, \rhotr_j) = 
 {}_2 F_1\left(\apsi_j + \cpsi_j,  \apsi_j + \cpsi_j, \apsi_j; \frac{\apsi_j^2\rhotr_j\psis_{jt}\psis_{j,t-1}}{(\apsi_j\psis_{jt} + \cpsi_j(1 - \rhotr_j))(\apsi_j\psis_{jt-1} + \cpsi_j(1 - \rhotr_j))}\right) \\
 &\times \left(\frac{\cpsi_j^2\left(\frac{\apsi_j}{\cpsi_j} + \frac{1}{\psis_{j,t-1}}\right)(1 - \rhotr_j)\psis_{j,t-1}}{\left(\cpsi_j + \frac{\apsi_j \psis_{jt}}{1 - \rhotr_j}\right)(\apsi_j\psis_{j,t-1} + \cpsi_j(1 - \rhotr_j))}\right)^{\apsi_j + \cpsi_j} \left(\frac{\apsi_j}{\cpsi_j(1 - \rhotr_j)}\right)^{\apsi_j} \frac{\psis_{jt}^{\apsi_j - 1}}{B(\apsi_j, \cpsi_j)},
\end{aligned} 
\end{equation}
where ${}_2 F_1(a, b, c; z)$ is the (Gaussian or ordinary) hypergeometric function 
\begin{equation} \label{hypergeo_def}
    { }_2 F_1(a, b , c ; z)=\sum_{n=0}^{\infty} \frac{(a)_n(b)_n}{(c)_n} \frac{z^n}{n !},
\end{equation}
and 
\SFS{$(q)_n = {\Gamfun{q+n}}/{\Gamfun{q}}$ }
are the rising factorials.
\end{theorem}

\begin{figure}[t!]
\vspace{-1em}
  \begin{center}
   \scalebox{.55}
   {\includegraphics{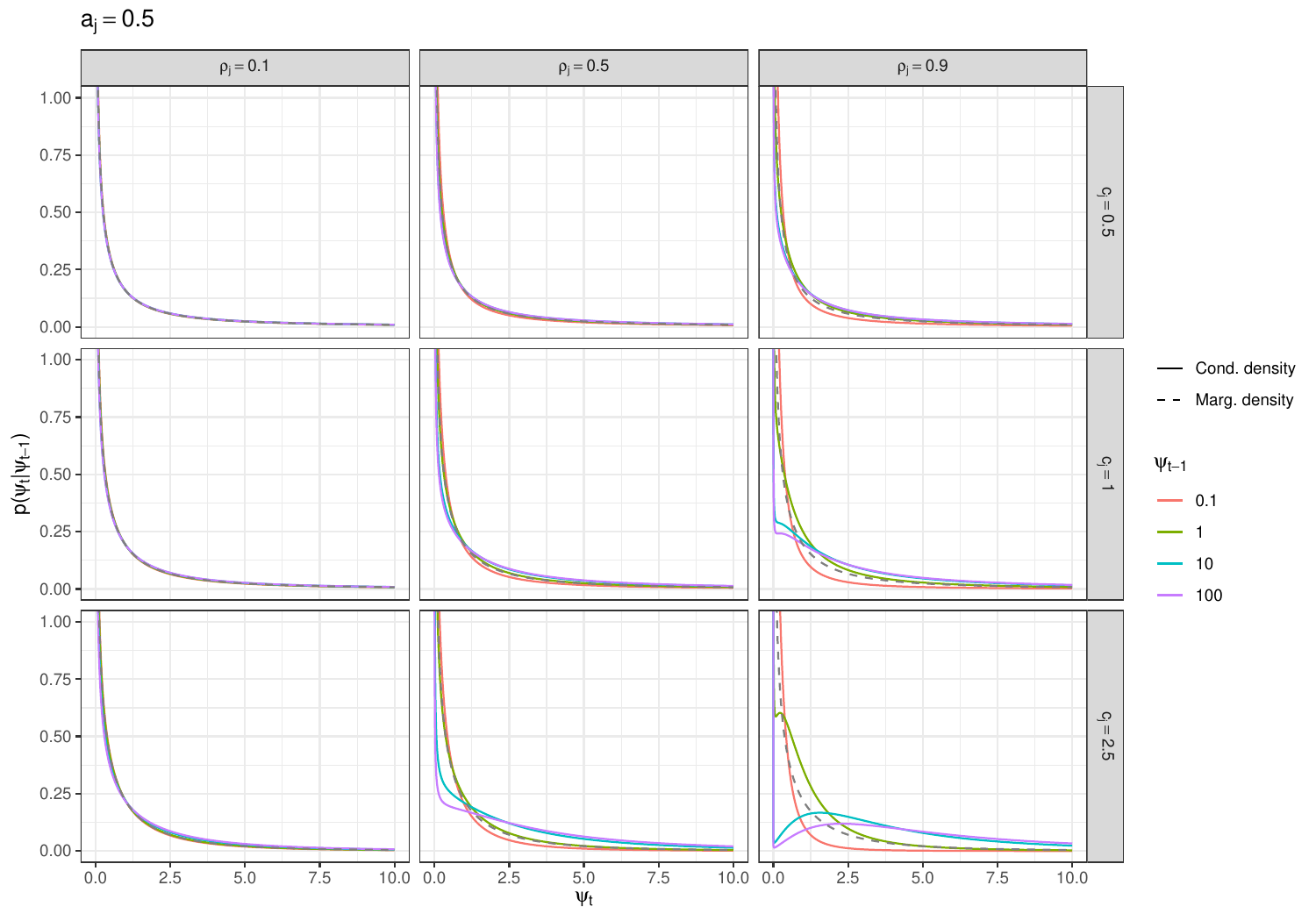}}
  \caption{\small Transition density $p(\psis_{jt} | \psis_{j,t-1}, \apsi_j, \cpsi_j, \rhotr_j)$ under the assumption that $\lambdac_{j,t-1}$ comes from the stationary distribution for $\apsi_j = 0.5$ and $\cpsi_j \in \{0.5, 1, 3\}$. Coloured lines represent transition densities for different $\psis_{j,t-1}$, while grey dashed lines represent the marginal density.}\label{fig_cond_dens}
  \end{center}
  \vspace{-1em}
\end{figure}

Theorem~\ref{theo_cond} is useful both for building understanding of the properties of the \SFSnew{DTG} prior, as well as for the MCMC algorithm presented in Section~\ref{compu}. Figure~\ref{fig_cond_dens} plots the transition density for various values of $\rhotr_j$, $\cpsi_j$ and $\psis_{j,t-1}$, highlighting the role both \SFSnew{model parameters} 
play in determining the dynamic shrinkage characteristics of the DTG \SFSnew{prior}. Starting with $\rhotr_j$, one can see that the degree to which the 
\SFS{conditional}
density deviates from the marginal density for large values of $\psis_{j,t-1}$ depends strongly on the value of $\rhotr_j$, with values closer to $1$ corresponding to stronger deviation. Whereas all densities in the left-most column, where $\rhotr_j = 0.1$, hew closely to the marginal distribution, there exists significant variation in the right-most column, where $\rhotr_j = 0.9$. Despite this, it is interesting to note that 
\SFS{fixing $\apsi_j$ at 0.5}
still causes all distributions to feature an infinite pole at the origin, irrespective of the value of $\psis_{j,t-1}$, with some distributions displaying multimodal behavior.  This effect of $\rhotr_j$ aligns with the idea of it being a parameter that controls dependence, as, the larger $\rhotr_j$ is, the more large values of $\psis_{j,t-1}$ increase the probability of observing large values of $\psis_{jt}$. 

Shifting focus to the effect of $\cpsi_j$ reveals that, as in the static triple gamma, $\cpsi_j$ controls the tails of the conditional density. Roughly speaking, smaller values of $\cpsi_j$ lead to heavier tails, an idea that is also reflected in the fact that the marginal distribution has no moments for $\cpsi_j < 0.5$. This idea is extended to $p(\psis_{jt} | \psis_{j,t-1}, \apsi_j, \cpsi_j, \rhotr_j)$ in Theorem~\ref{expect_cond}, with the proof again found in Appendix~\ref{proofs}.
\begin{theorem}\label{expect_cond}
For $\psis_{jt} | \psis_{j,t-1}, \apsi_j, \cpsi_j, \rhotr_j$ with density defined in Theorem~\ref{theo_cond} and  $\apsi_j > 0$,  $\cpsi_j > 1$, $0 < \rhotr_j < 1$, and $\lambdac_{j,t-1} \sim \Gammad{\apsi_j, \frac{\apsi_j}{\cpsi_j}} $, the following holds:
\begin{equation} \label{expec_cond}
\mathbb E \left[ \psis_{jt} | \psis_{j,t-1}, \apsi_j, \cpsi_j > 1, \rhotr_j \right] = (1 - \rhotr_j)\frac{\cpsi_j}{\cpsi_j - 1} + \rhotr_j \frac{\psis_{j,t-1}\cpsi_j(\apsi_j + \cpsi_j)}{(\cpsi_j + \apsi_j\psis_{j,t-1})(\cpsi_j - 1)}.
\end{equation}
\end{theorem}
The implications of Theorem~\ref{expect_cond} align neatly with the intuition gained from Figure~\ref{fig_cond_dens}. The expectation is a weighted average of the expected value 
\SFS{$\cpsi_j/(\cpsi_j - 1)$} 
of the marginal $\Fd{2\apsi_j, 2\cpsi_j}$-distribution 
and an expression depending non-linearly on $\psis_{j,t-1}$. Therefore, it deviates further from the marginal expectation the closer $\rhotr_j$ is to $1$. The tail controlling behavior of $\cpsi_j$ becomes apparent if one lets $\cpsi_j$ approach $1$ from above, as 
$\lim_{\cpsi_j \downarrow 1} \mathbb E \left[ \psis_{jt} | \psis_{j,t-1}, \apsi_j, \cpsi_j, \rhotr_j \right] = \infty$.
On the other hand, the fact that values of $\cpsi_j > 1$ imply heavier regularisation of the tails is also evident, as the expectation, while monotonically increasing in $\psis_{j,t-1}$, is not unbounded. Specifically,
\begin{align*}
        \lim_{\psis_{j,t-1} \rightarrow \infty} \mathbb E \left[ \psis_{jt} | \psis_{j,t-1}, \apsi_j, \cpsi_j > 1, \rhotr_j \right] =  \frac{\cpsi_j(\apsi_j + \cpsi_j\rhotr_j)}{\apsi_j(\cpsi_j - 1)}.
\end{align*} 

\begin{figure}[t!]
\vspace{-1em}
  \begin{center}
   \scalebox{.55}
   {\includegraphics{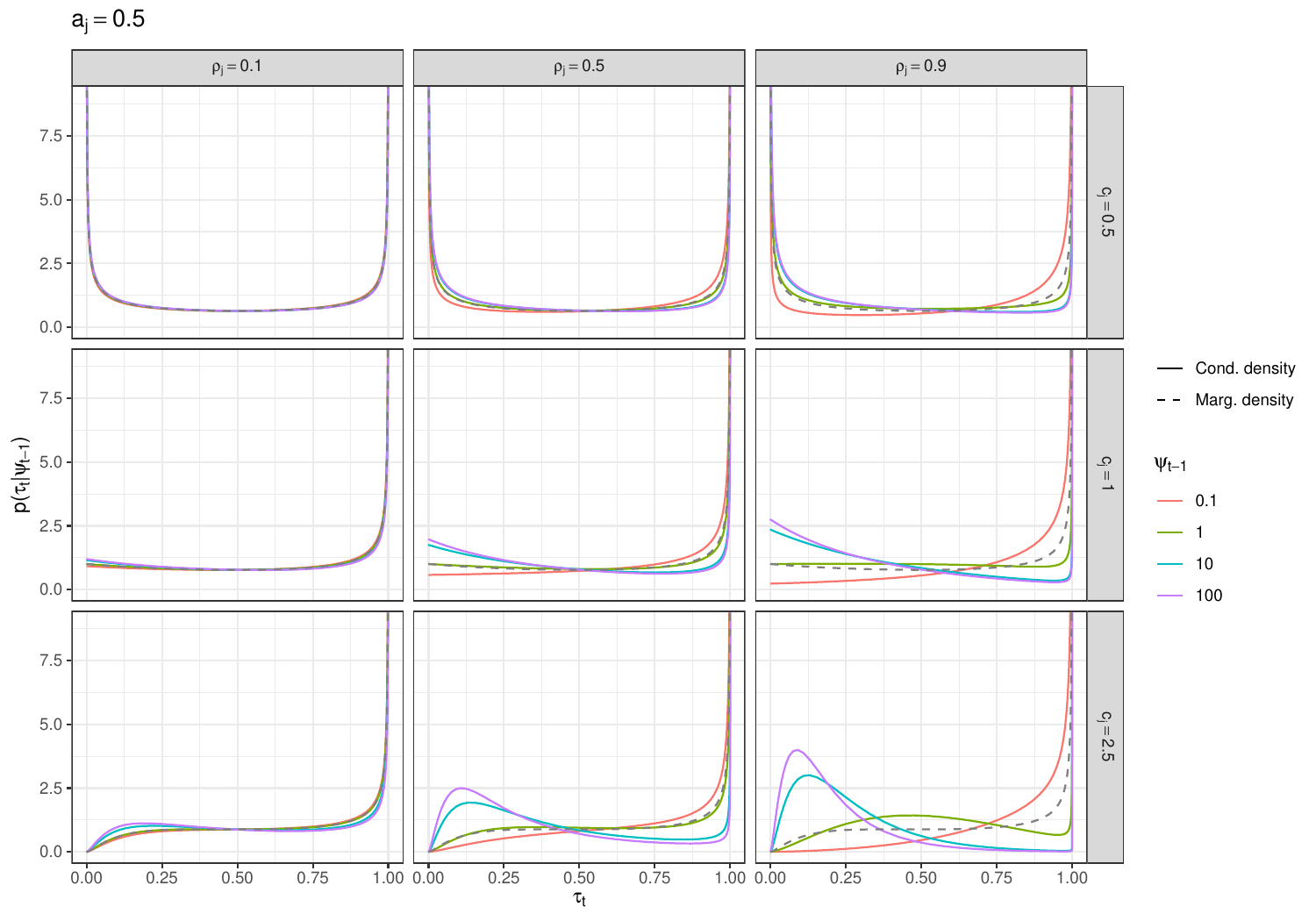}}
  \caption{\small Dynamic shrinkage profiles $p(\tau_{jt} | \psis_{j,t-1}, \apsi_j, \cpsi_j, \rhotr_j)$ for $\apsi_j = 0.5$ and $\cpsi_j \in \{0.5, 1, 2.5\}$. Coloured lines represent shrinkage profiles for different $\psis_{j,t-1}$, while grey dashed lines represent the marginal shrinkage profile.}\label{fig_shri_prof}
  \end{center}
  \vspace{-1em}
\end{figure}

\noindent 
Another benefit of Theorem~\ref{theo_cond} is that it allows for the derivation of conditional shrinkage profiles, in the spirit of \cite{car-etal:hor}. Letting $\kfQc_j = 1$ for illustrative purposes, one can re-parameterize the conditional density of $\kfwc_{jt}$ in the following way, 
\SFS{where $\psis_{jt} =1 /{\tau_{jt}} - 1 $:}
\begin{equation} \label{wshrink}
    \kfwc_{jt}|\tau_{jt} \sim \Normal{0, 1 /{\tau_{jt}} - 1}.
\end{equation}
Using the law of transformation of densities 
\SFS{for $\tau_{jt} = \frac{1}{1 + \psis_{jt}}$}
yields the following 
corollary.

\begin{corollary}
Let $\tau_{jt} = \frac{1}{1 + \psis_{jt}}$, where $\psis_{jt} | \psis_{j,t-1}, \apsi_j, \cpsi_j, \rhotr_j$ is a random variable with density defined in Theorem~\ref{theo_cond} and $\apsi_j > 0$,  $\cpsi_j > 0$, $0 < \rhotr_j < 1$, and $\lambdac_{j,t-1} \sim \Gammad{\apsi_j, \frac{\apsi_j}{\cpsi_j}}$, then
 \begin{align}
    p(\tau_{jt}|\psis_{j,t-1}, \apsi_j, \cpsi_j, \rhotr_j) = p_{\psis_{jt}|\psis_{j,t-1}}\left(\frac{1}{\tau_{jt}} - 1 , \apsi_j, \cpsi_j, \rhotr_j \right)\frac{1}{\tau_{jt}^2},
\end{align}
where $p_{\psis_{jt}|\psis_{j,t-1}}\left(x,\apsi_j, \cpsi_j, \rhotr_j \right)$ 
is shorthand for the transition density defined in \eqref{psi_cond}.
\end{corollary}
 $\tau_{jt}$ is confined to the range $(0, 1)$ and characterizes the shrinkage behavior, with $\tau_{jt}$ close to $1$ implying "total" shrinkage, as the variance of $\kfwc_{jt}$ then approaches 0 and $\tau_{jt}$ close to $0$ implying the opposite. Figure~\ref{fig_shri_prof} graphs the dynamic shrinkage profiles for various values of $\rhotr_j$, $\cpsi_j$ and $\psis_{j,t-1}$. The role of $\rhotr_j$ as the parameter 
 \SFS{driving}
 the extent to which the conditional distribution deviates from the marginal distribution is again 
  apparent. Similarly, $\cpsi_j$'s importance for the shape of the tails of the distribution is reinforced. In the top row, where $\cpsi_j = 0.5$, the result is a dynamic horseshoe, with infinite poles in both corners, irrespective of the magnitude of $\psis_{j,t-1}$, with larger values of $\psis_{j,t-1}$ allocating more prior mass toward $0$ and vice versa.
 In the bottom two rows, where 
 $\cpsi_j \geq 1$, the infinite pole in the left corner disappears, implying some regularization in the tails, even for large values of $\psis_{j,t-1}$. This lines up nicely with the results derived from Theorem~\ref{expect_cond}, as the expected value 
 of the conditional density only exists for $\cpsi_j > 1$.

\section{MCMC estimation} \label{compu}

Performing inference in such a highly parameterized model is challenging both from a theoretical as well as a computational perspective. These challenges stem almost exclusively from the latent non-Gaussian processes
$\psis_{jt}$'s and the associated hyperpriors, which becomes evident when noting that the model simplifies to a linear Gaussian state space model when conditioning on the $\psis_{jt}$'s. As this is a well-understood class of models, we can fall back on algorithms that are fairly standard in the literature for the portions of the MCMC algorithm that deal with a conditionally Gaussian state space model. This includes, for example, the simulation smoother of  \cite{mcc-etal:sim} or the ancillarity-sufficiency interweaving strategy (ASIS) of 
\SFSnew{\cite{bit-fru:ach}} 
to increase the sampling efficiency of $\kfQc_1, \dots, \kfQc_d$. Therefore, we focus on the parts of the algorithm that are non-standard here, by conditioning on the innovations $\kfwc_{jt} = \betaci{jt} - \betaci{j,t-1}$. We give a full overview of the MCMC scheme in Appendix~\ref{sec_details}. Readers interested in estimating TVP models within the DTG framework can find implementations of the algorithms discussed here in the R package \texttt{shrinkTVP}, available on CRAN.

\paragraph*{Sampling under \SFSnew{exchangeable triple gamma} innovations}

To give an intuitive understanding of the MCMC sampler, we begin by discussing how to draw samples from 
a simple model, characterized by 
\SFSnew{a priori exchangeable} triple gamma innovations, i.e. $\kfwc^2_{jt} |\kfQc_j, \apsi_j , \cpsi_j \mathop{\sim}\limits^{\mathrm{iid}} TG\left(\apsi_j, \cpsi_j, 2/\kfQc_j\right)$. This is a special case of the DTG that results when $\rhotr_j = 0$, which causes all $d$ sets of latent variables $\kappaPG_{j1}, \dots, \kappaPG_{jT}$
in (\ref{kappa}) to be equal to 0, while
the latent variables $\lambdac_{j1}, \ldots, \lambdac_{jT}$ in (\ref{lambda}) arise independently from 
$\lambdac_{jt}|\apsi_j, \cpsi_j \mathop{\sim}\limits^{\mathrm{iid}}  \Gammad{\apsi_j, \frac{\apsi_j}{\cpsi_j} }$ \SFSnew{as in \eqref{rephor}}.

Here, one can employ either the sampler proposed in \cite{cad-etal:tri} or, based on representation \eqref{rephor}, utilize a generalized and innovative version of the sampler introduced for the horseshoe prior by \cite{mak-sch:sim}. Applying standard derivations for conditionally conjugate densities involving the inverse gamma and gamma distributions yields:
\begin{align}
\psis_{jt}|\kfwc_{jt}, \lambdac_{jt}, \kfQc_j, \cpsi_j&\sim \Gammainv{\cpsi_j + \frac 1 2, \frac{\kfwc^2_{jt}}{2\kfQc_j} + \lambdac_{jt}}, \label{psi_cond_post}\\
\lambdac_{jt}|\psis_{jt},\apsi_j, \cpsi_j &\sim \Gammad{\apsi_j + \cpsi_j, \frac{\apsi_j}{\cpsi_j} + \frac{1}{\psis_{jt}}}, \label{lambda_cond_post}
\end{align}
for $t = 1, \dots, T$ and $j = 1, \dots, d$. This result is valuable in its own right, offering a method for sampling from the posterior of a model under the triple gamma prior without relying on the generalized inverse Gaussian distribution (GIG) as in \cite{cad-etal:tri}. As the GIG is not as widely implemented in programming languages compared to the gamma and inverse gamma distributions, this outcome facilitates the broader application of the triple gamma prior.

\paragraph*{Sampling for fixed \SFS{\texorpdfstring{$\rhotr_j>0$}{rho j >0}}}

Moving from the case where the innovations are exchangeable triple gamma, the next simplest case is one where dependence is modeled via a fixed 
\SFS{$\rhotr_j >0$}.
Naturally, this additionally necessitates the sampling of the latent variables $\kappaPG_{j1}, \dots, \kappaPG_{jT}$. Here, the hierarchical structure allows a 
straightforward Gibbs sampler to be implemented to sample from the posterior of 
\SFS{all unknowns}.
The conditional density \eqref{psi_cond_post} for $\psis_{jt}$ does not require any modification, as 
conditional on $\lambdac_{jt}$, it is independent of $\kappaPG_{j1}, \dots, \kappaPG_{jT}$. The full conditional posterior for $\lambdac_{jt}$ can, once again, be found through fairly standard derivations based on gamma and inverse gamma densities:\footnote{In several of the sampling steps discussed in this section, adjustments are necessary for the cases where $t \in \{0, 1, T\}$. To maintain brevity, these modifications can be found in Appendix~\ref{sec_details}.} 
\begin{eqnarray} \label{renhgpost}
\lambdac_{jt} | \psis_{jt} ,  \kappaPG_{jt},  \kappaPG_{j,t+1}, \apsi_j, \cpsi_j, \rhotr _j
\sim \Gammad{\apsi_j + \cpsi_j + \kappaPG_{jt} + \kappaPG_{j,t+1},
\frac{\apsi_j}{\cpsi_j}\frac{1 + \rhotr_j}{1- \rhotr_j } + \frac{1}{\psis_{jt} }},
\end{eqnarray}
where the 
case $\rhotr_j=0$ leads back to (\ref{lambda_cond_post}). The  conditional posterior
$\kappaPG_{jt} | \lambdac_{jt}, \lambdac_{j,t-1}, \apsi_j, \cpsi_j, \rhotr _j  $  follows a discrete distribution,  independently of all other state variables, with weights given by:
\begin{eqnarray} \label{kappaden}
  \Prob{\kappaPG_{jt}=k| \lambdac_{jt}, \lambdac_{j,t-1}, \apsi_j, \cpsi_j, \rhotr _j} \propto 
\SFS{ \frac{(\lambdac_{jt} \lambdac_{j,t-1}) ^k}{\Gamfun{\apsi_j + k} \Gamfun{k+1}} 
  \left( \frac{\apsi_j \sqrt{\rhotr_j}}{\cpsi_j(1- \rhotr_j )} \right) ^{2k} .}
\end{eqnarray}
Sampling in this fully conditional setup works well, with good mixing and computations that are able to be performed fairly quickly. However, fixing $\rhotr_j$ a priori is a fairly strong assumption, particularly because it has a large influence on the properties of the DTG (see Section~\ref{sec_dtg}).  

\paragraph*{Sampling with unknown \texorpdfstring{$\rhotr_j$}{rho j}}

As opposed to the sampling steps for $\psis_{jt}$, $\lambdac_{jt}$ and $\kappaPG_{jt}$, the full conditional posterior of $\rhotr_j$ is not available in closed form, necessitating the use of a Metropolis-Hastings-within-Gibbs step. Implementing such a step within a fully conditional Gibbs loop in a straightforward fashion leads to poor mixing. 
To alleviate this issue, we make use of representations of the hierarchical structure with different sets of random variables marginalized out. 
For the Metropolis-Hastings (MH) step for $\rhotr_j$, we construct an approximate likelihood using Theorem~\ref{theo_cond}:
\begin{equation} \label{approx_lik}
    p(\psi_{j1}, \dots, \psi_{jT}|\rhotr_j, \apsi_j, \cpsi_j) \approx p(\psis_{j1} | \apsi_j, \cpsi_j, \rhotr_j) \prod_{t = 2}^T p(\psis_{jt} | \psis_{j,t-1}, \apsi_j, \cpsi_j, \rhotr_j),
\end{equation}
where $p(\psis_{jt} | \psis_{j,t-1}, \apsi_j, \cpsi_j, \rhotr_j)$ is the density defined in \eqref{psi_cond}. This, combined with the prior density defined in Section~\ref{prior_choices}, gives us the posterior up to an unknown constant, allowing the construction of an MH step to generate samples from the conditional posterior $p(\rhotr_j| \apsi_j, \cpsi_j,\psi_{j1}, \dots, \psi_{jT}) $. 
The important thing to note here is that this approximate likelihood is marginalized \SFSnew{both} w.r.t. $\lambdac_{j0}, \dots, \lambdac_{jT}$ and $\kappaPG_{j1},\dots, \kappaPG_{jT}$. Thus, to preserve the stationary distribution of the Markov chain, one needs to sample either $\lambdac_{j0}, \dots, \lambdac_{jT}$ or $\kappaPG_{j1},\dots, \kappaPG_{jT}$ from \SFSnew{posterior} densities that are marginalized w.r.t. to the respective other \citep{van-par:par}. 

\begin{figure}[t!]
\vspace{-1em}
\centering
\scalebox{0.75} {
\begin{tikzpicture}[node distance = {5mm}, 
                    main/.style = {draw, circle},  
                    minimum size = 1.4cm] 
                    
\node[main] (1) {$\kfwc_{j,t-1}$};
\node[main, right = of 1] (2) {$\kfwc_{jt}$};
\node[main, right = of 2] (3) {$\kfwc_{j,t+1}$};

\node[main] (4) [below = of 1] {$\psis_{j,t-1}$};
\node[main] (5) [below = of 2] {$\psis_{jt}$};
\node[main] (6) [below = of 3] {$\psis_{j,t+1}$};

\node[main] (7) [below = of 4] {$\kappaPG_{j,t-1}$};
\node[main] (8) [below = of 5] {$\kappaPG_{jt}$};
\node[main] (9) [below = of 6] {$\kappaPG_{j,t+1}$};

\node[main, draw = none] (10) [left = of 4] {};
\node[main, draw = none] (11) [left = of 7] {};
\node[main, draw = none] (12) [right = of 9] {};

\draw[->] (4) -- (1);
\draw[->] (5) -- (2);
\draw[->] (6) -- (3);

\draw[->] (10) -- (7);
\draw[->] (11) -- (7);
\draw[->] (7) -- (4);

\draw[->] (4) -- (8);
\draw[->] (7) -- (8);
\draw[->] (8) -- (5);

\draw[->] (5) -- (9);
\draw[->] (8) -- (9);
\draw[->] (9) -- (6);

\draw[->] (6) -- (12);
\draw[->] (9) -- (12);

\end{tikzpicture} 
}
\caption{\small Interdependencies of the latent variables of the dynamic triple gamma prior represented as a directed graph, in the alternate state space model defined by 
\SFS{\eqref{psijtpostD_pap} -- \eqref{KjtpostD_pap}.}
}
\label{tg_graph_alt}
\vspace{-1em}
\end{figure}
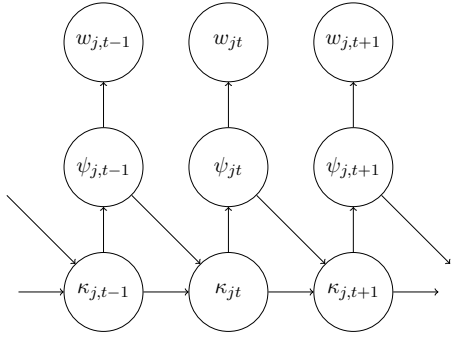

It is possible to construct a state space model with the same marginal properties as the one introduced in Section~\ref{sec_dtg}, albeit with $\SFSnewRed{\lambdac_{j0}}, \dots, \lambda_{jT}$ marginalized out. It is characterized by a generalized beta prime transition density for $\psis_{jt} | \kappaPG_{jt}, \apsi_j, \cpsi_j, \rhotr _j$, specifically 
\begin{eqnarray} \label{psijtpostD_pap}
\psis_{jt} | \kappaPG_{jt}, \apsi_j, \cpsi_j, \rhotr _j \sim \GenBetapr{\apsi_j + \kappaPG_{jt}, \cpsi_j, 1, \frac{\cpsi_j(1 - \rhotr_j )}{\apsi_j}},
\end{eqnarray}
coupled with the following negative binomial transition density for $\kappaPG_{jt}  | \kappaPG_{j,t-1},  \psis_{j,t-1} , \rhotr_j, \apsi_j, \cpsi_j$:
\begin{eqnarray} \label{KjtpostD_pap}
\kappaPG_{jt}  | \kappaPG_{j,t-1},  \psis_{j,t-1} , \rhotr_j, \apsi_j, \cpsi_j &\sim &
\NegBin{\apsi_j + \cpsi_j + \kappaPG_{j,t-1}, \piNB _{j,t-1}}, \\  \nonumber
 \piNB_{j,t-1} & = &
 \frac{\apsi_j\psis_{j,t-1} + \cpsi_j (1- \rhotr_j) }{(1+\rhotr_j) \apsi_j\psis_{j,t-1} + \cpsi_j (1- \rhotr_j) },
\end{eqnarray}
with the initial value coming from the stationary distribution of the process governing $\kappaPG_{jt}$, i.e. $\kappaPG_{j1} | \apsi_j, \rhotr_j \sim \NegBin{\apsi_j , 1-\rhotr_j }.$
A representation of the interdependencies of this partially marginalized state space model as a directed graph can be found in Figure~\ref{tg_graph_alt}. Details on the derivations of transition densities and the stationary distribution can be found in Appendix~\ref{state_spaces}.

In this alternate state space representation, we can now sample $\kappaPG_{j1}, \dots, \kappaPG_{jT}$ conditional on $\psis_{j1}, \dots, \psis_{jT}$ but marginalized w.r.t. $\lambdac_{j0}, \dots, \lambdac_{jT}$, \SFS{as required.}
More specifically
\begin{eqnarray} 
   \label{kjtt_pap}
&&  p( \kappaPG_{jt}  | \kappaPG_{j,t-1}, \kappaPG_{j,t+1}, \psis_{jt}, \psis_{j,t-1},\rhotr_j, \apsi_j, \cpsi_j )  \\
& \propto & \frac{\Gamfun{\apsi_j + \cpsi_j + \kappaPG_{j,t-1} + \kappaPG_{jt}}\Gamfun{\apsi_j + \cpsi_j + \kappaPG_{jt} + \kappaPG_{j,t+1}}}
{ \kappaPG_{jt} ! \Gamfun{\apsi_j  + \kappaPG_{jt}}}
     \left( z^\psi_{jt}
     \right) ^ {\kappaPG_{jt}}, \nonumber    \end{eqnarray}
with  
$z^\psi_{jt} = (1-\piNB _{j,t-1})\piNB_{jt}\apsi_j \psis_{jt}/(\apsi_j \psis_{jt}  + \cpsi_j (1-\rhotr _j) )$ and  $\piNB_{j,t-1}$ and $\piNB_{jt}$ defined as in~\eqref{KjtpostD_pap}. 
Further, it can be shown that the normalizing constant of this distribution is available in closed form. 
From the definition of the hypergeometric function presented in \eqref{hypergeo_def}, 
it follows immediately that following infinite series can be represented by a hypergeometric function:
\begin{eqnarray*}
\sum_{k=0}^\infty 
\frac{\Gamfun{a+k} \Gamfun{b+k}}{\Gamfun{c+k}}
 \frac{z^k}{k!} = \frac{\Gamfun{a} \Gamfun{b}}{\Gamfun{c}} {}_2 F_1(a, b; c; z) .
\end{eqnarray*}
Therefore, the normalizing constant of \eqref{kjtt_pap} is equal to
\begin{eqnarray} \label{normkap_pap}
 \sum_{\kappa_{jt}=0}^\infty  p( \kappaPG_{jt}  | \kappaPG_{j,t-1}, \kappaPG_{j,t+1}, \psis_{jt}, \psis_{j,t-1},\rhotr_j, \apsi_j, \cpsi_j )
 = \frac{\Gamfun{a^\kappa_{jt}} \Gamfun{b^\kappa_{jt}}}{\Gamfun{\apsi_j}} {}_2 F_1\left(a^\kappa_{jt}, b^\kappa_{jt}; \apsi_j; 
z^\psi_{jt}
 \right) ,
\end{eqnarray} 
with $a^\kappa_{jt} = \apsi_j + \cpsi_j + \kappaPG_{j,t-1}$ and $b^\kappa_{jt} = \apsi_j + \cpsi_j + \kappaPG_{j,t+1}$. 
 
Based on the normalized weights defined by \eqref{kjtt_pap} and \eqref{normkap_pap}, two main options present themselves to sample $\kappaPG_{jt}$. One could adapt Algorithm~11.5 in \citet[p.342]{fru:book}, to derive a forward-filtering-backward-sampling (FFBS) algorithm. 
This re-frames $\kappaPG_{jt}$ as a hidden Markov chain with $\Kmax$ 
states, with the 
\SFS{rows of the transition matrix}
given by the normalized version of \eqref{kjtt_pap}. The 
attractiveness of this approach lies in the ability to 
\SFS{jointly draw}
\SFSnew{the entire path}
$\kappaPG_{j1}, \dots, \kappaPG_{jT}$, thereby improving mixing. The downside 
is that 
the computation of all $\Kmax^2\times (T-1)$ transition matrices 
\SFS{is required} 
before sampling can be performed, which becomes fairly computationally demanding for even moderately sized problems. 

An alternate approach consists of sampling $\kappaPG_{jt}$ directly using inverse transform sampling 
\SFSnew{for the normalized version of \eqref{kjtt_pap}}. 
The probabilities have a convenient recursive structure (see Appendix~\ref{sec_details}), allowing for efficient computation, as only the probabilities up to the current realization of $\kappaPG_{jt}$ need to be evaluated.
Thus, particularly when many values of $\kappaPG_{jt}$ are zero or near zero (as is the case with heavy shrinkage), this method can be executed orders of magnitude faster than FFBS. Further, we found that mixing was fairly good when using this method, despite it being a single move sampler. As a result of its superior speed and generally favorable mixing, we advocate for adopting this method to sample $\kappaPG_{jt}$.

In summary, the proposed sampler 
\SFS{alleviates}
poor mixing for $\rhotr_j$ by using the approximate likelihood \eqref{approx_lik} marginalized w.r.t. $\SFSnewRed{\lambdac_{j0}}, \dots, \lambdac_{jT}$ and $\kappaPG_{jt}, \dots, \kappaPG_{jT}$ to construct a collapsed MH-within-Gibbs step for $\rhotr_j$. 
\SFSnew{A partially marginalized state space model is  utilized to}
sample $\kappaPG_{j1}, \dots, \kappaPG_{jT}$ conditional on $\psis_{j1}, \dots, \psis_{jT}$ but marginalized w.r.t. $\lambdac_{j0}, \dots, \lambdac_{jT}$. This is done by applying inverse transform sampling to the normalized version of \eqref{kjtt_pap}. Next, one can sample $\SFSnewRed{\lambdac_{j0}}, \dots, \lambdac_{jT}$ 
\SFS{from \eqref{renhgpost}}
as in the fixed $\rhotr_j$ case.
Finally, conditional on $\SFSnewRed{\lambdac_{j0}}, \dots, \lambdac_{jT}$, realizations of 
$\psis_{j1}, \dots, \psis_{jT}$ are sampled from \eqref{psi_cond_post} as for exchangeable innovations.  It is important to note that the sampling order is important when marginalizing out random variables, as the stationary distribution of the Markov chain is not invariant to permutations of the sampling steps.

\section{Application to the EURO STOXX 50 Index} \label{appl}

The Cholesky SV-model developed by \cite{lop-etal:par} and further refined by \cite{bit-fru:ach} provides a method for modeling sparse, time-varying variance-covariance matrices $\bm \Sigma_t$, for $t = 1, \dots, T$, of an $M$-dimensional multivariate time series that follows a conditionally zero mean normal distribution, i.e. $\bm y_t | \bm \Sigma_t \sim \mathcal N_M\left(\bm 0, \bm \Sigma_t\right)$. The key ingredient is the decomposition $\bm \Sigma_t = \bm\Lambda_t\bm D_t \bm \Lambda_t'$, where $\bm \Lambda_t$ is lower unitriangular and $\bm D_t$ is a diagonal matrix. As a consequence, $\bm\Lambda_t^{-1}\bm y_t\sim\mathcal N_M\left(\bm 0, \bm D_t\right)$, which, letting the elements of $\bm\Lambda_t^{-1}$ be denoted as 
\SFSred{$\gamma_{mj,t}$ for $j < m$},
can be expressed as 
$$
{\small \left[\begin{array}{ccccc}
1 & \ldots & & & 0 \\
\gamma_{21, t} & 1 & & & 0 \\
& & \ddots & & 0 \\
\vdots & & & 1 & 0 \\
\gamma_{M 1, t} & \gamma_{M 2, t} & \ldots & \gamma_{M, M-1, t} & 1
\end{array}\right]\left[\begin{array}{c}
y_{1 t} \\
y_{2 t} \\
\vdots \\
y_{M t}
\end{array}\right] \sim \mathcal{N}_M\left(\mathbf{0}, \mathbf{D}_t\right).}
$$
As $\bm D_t$ is a diagonal matrix, $M$ independent TVP models result:
{\small  

\vspace*{-12mm}
\begin{eqnarray*}
    && y_{1t} = \varepsilon_{1t}, \quad \varepsilon_{1t} \sim \Normal{0, \sigma^2_{1t}}, \\
    && y_{2t} = -\gamma_{21,t} y_{1t} + \varepsilon_{2t}, \quad \varepsilon_{2t} \sim \Normal{0, \sigma^2_{2t}}, 
    \\
    && \qquad\qquad\vdots \\
    && y_{Mt} = -\gamma_{M1,t} y_{1t}- \dots -\gamma_{M,M-1,t} y_{M-1,t} + \varepsilon_{Mt}, \quad \varepsilon_{Mt} \sim \Normal{0, \sigma^2_{Mt}}.
\end{eqnarray*}}
To further allow for conditional heteroscedasticity, one can let the 
\SFSnewRed{$\sigma^2_{mt} = e^{h_{mt}}$ }
follow a stochastic volatility (SV) law of motion,
\SFSnew{for $m=1, \ldots, M$, where:} 
\begin{eqnarray} \label{svht}
\SFSnewRed{h_{mt} | h_{m,t-1}, \mu_m, \phi_m, \sigma_{\eta,m}^2 \sim \Normal{\mu_m + \phi_m(h_{m,t-1}-\mu_m),\sigma_{\eta,m}^2}.}
\end{eqnarray}
For the priors on the \SFSnew{model parameters} 
of the SV process, we follow \cite{kas-fru:anc}, as these are well-established in the literature, \SFSnew{see Appendix~\ref{sec_prior_details} for details.}

The equations above can now be estimated as a series of independent TVP regression models with univariate outcomes $y_{mt}$, where the $\gamma_{mj,t}$ (and hence also the elements of $\bm\Lambda_{t}^{-1}$) can be recovered from the regression coefficients as $-\beta_{mj,t}$. The idiosyncratic variance matrix $\bm D_t = \Diag{e^{h_{1t}}, \dots, e^{h_{Mt}}}$ can be recovered from 
the log volatilities of the individual equations, with the exception of $e^{h_{1t}}$, for which, due to the absence of any regressors, an SV model has to be estimated separately. Together, the draws from $\bm\Lambda_{t}^{-1}$ and $\bm D_t$ define the posterior draws of $\bm \Sigma_t$, binding the draws from the individual equations together. Placing a dynamic triple gamma prior on the innovations of the regression coefficients $\beta_{mj,t}$ allows for covariance modeling that is highly sparse, due to the strong shrinkage imposed by the prior, while still allowing for adaptation to locally different (co)variances, thanks to the dependence induced by the DTG.

\subsection{Time-Varying Covariances of Returns} \label{sec_tv_appl}

\begin{figure}[t!]
\vspace{-1em}
  \begin{center}
   \scalebox{0.6}
   {\includegraphics{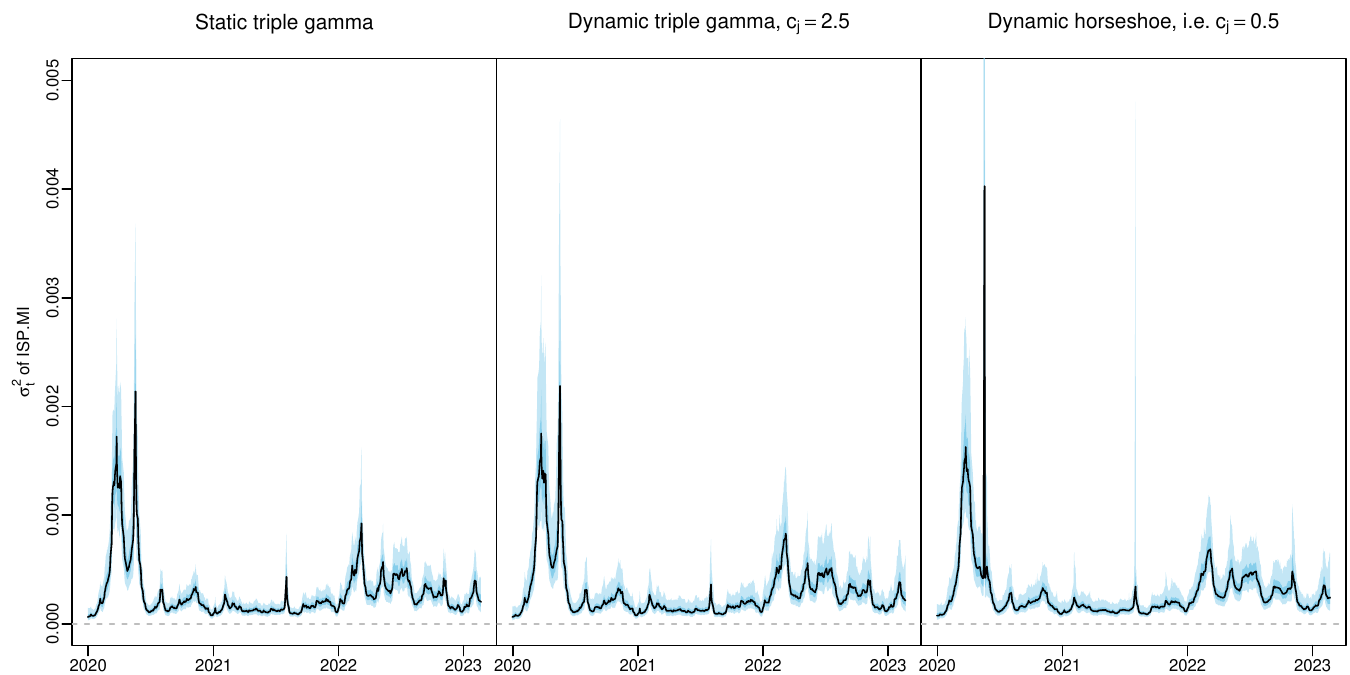}}
  \caption{\small 
  \SFS{Time-varying} variance of returns of Intesa Sanpaolo in a Cholesky SV-model under the static triple gamma prior, dynamic triple gamma and dynamic horseshoe. The black line represents the posterior median, while the blue shaded regions represent the point-wise 90\% and 50\% credible intervals.\label{fig_ISP}}
  \end{center}
  \vspace{-1em}
\end{figure}

\begin{figure}[t!]
\vspace{-1em}
  \begin{center}
   \scalebox{0.6}
   {\includegraphics{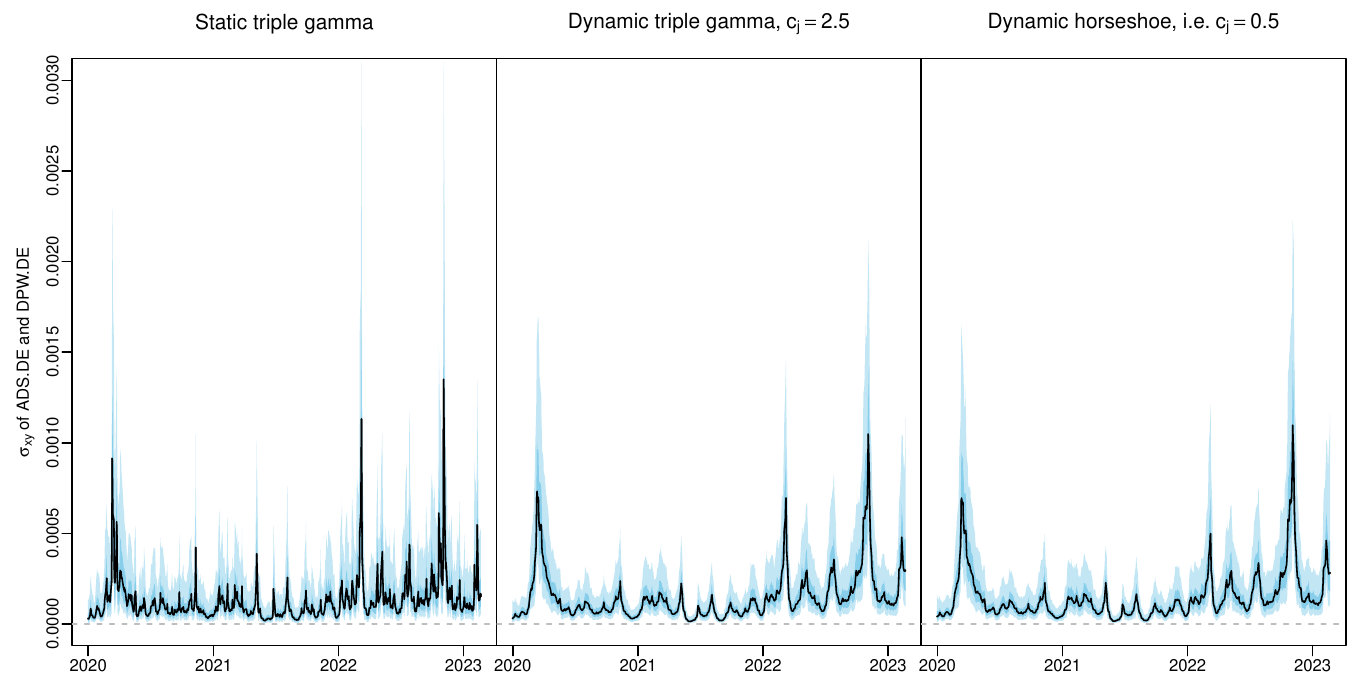}}
  \caption{\small 
  \SFS{Time-varying} covariance of returns of Adidas and Deutsche Post 
  under the static triple gamma prior, dynamic triple gamma and dynamic horseshoe. The black line represents the posterior median, while the blue shaded regions represent the point-wise 90\% and 50\% credible intervals.  \label{fig_DPW_ADS}}
  \end{center}
  \vspace{-1em}
\end{figure}

We apply the Cholesky SV-model equipped with the dynamic triple gamma prior to a data set of 45 out of 50 of the returns on the companies represented in the EURO STOXX 50 index (with 5 dropped due to data availability issues, see Appendix~\ref{data} for a more detailed description). The order of the data set is alphabetical and it spans 810 time points from the 2nd of January, 2020 to the 21st of February, 2023. Due to the COVID-19 pandemic and the ensuing volatility in financial markets, this represents a particularly challenging data set for econometric modeling. 
$\apsi_j$ is set to 0.5, as we want to induce heavy shrinkage around the origin, and we test two different choices of $\cpsi_j$, namely $\cpsi_j = 0.5$ and $\cpsi_j = 2.5$, as the former implies a marginal horseshoe prior for the innovations with no moments and very heavy tails, while the latter allows for some regularization of the tails.
In both setups, we use the GB1 prior for $\rhotr_j$ defined in \eqref{rho_prior}, with hyperparameters $a_\rho = 1$, $b_\rho = 0.95$, $\alpha_\rho = 0.5$, and $\beta_\rho = 0.5$. This results in a prior that has a horseshoe shape ranging from 0 to 0.95, incorporating the prior information that we either expect a state to display very little dependence or a large amount of dependence, but not a middling amount. For the global shrinkage parameters of the innovations as well as the means of the initial values we employ the triple gamma prior distributions $\betaci{j}^2 \sim TG(a^\tau, c^\tau, \lambda_B^2)$ and $\kfQc_j \sim TG(a^\xi, c^\xi, \kappaBj{B}^2)$, under the hyperpriors defined in 
\SFSnew{Appendix~\ref{sec_prior_details}}
with hyperparameters $\alpha_{a^\xi} = \alpha_{a^\tau} = \alpha_{c^\xi} = \alpha_{c^\tau} = 5$, $\beta_{a^\xi} = \beta_{a^\tau} = 10$ and $\beta_{c^\xi} = \beta_{c^\tau} = 2$. For the 
\SFSnew{model parameter of the SV} equations, we use the default choices of \cite{kna-etal-new:shr_TVP}. For comparison, we also estimate a Cholesky SV-model under a static 
triple gamma prior, using the default hyperparameters laid out in \cite{knaus2021shrinkage} and under the dynamic shrinkage process of \cite{kow-etal:dyn}, using the default hyperparameter values suggested by the authors. Each model was run for 100~000 iterations, with a burn-in of 20~000 and a thinning factor of 100.

\subsection{Results}

Generally speaking, most elements of the estimated time varying variance-covariance matrix $\bm\Sigma_t$ are very close to zero for large stretches of time. Notable exceptions are early 2020, which correlates with the onset of the COVID-19 pandemic, early 2022, which coincides with the start of the full-scale Russo-Ukrainian War, and late 2022, where the energy crisis lead to increased volatility in Europe. While all three prior specifications result in posterior estimates with this property, there exist two notable differences. First, the dynamic horseshoe displays large, sudden jumps followed by periods of relative calm, showcasing the flexibility of the dynamic triple gamma prior approach, as well as the heavy tails of the horseshoe prior. An example of this behavior can be seen in the third panel of Figure~\ref{fig_ISP}, in the first four months of 2020. Second, the under-shrinking behavior of the static triple gamma approach becomes apparent in some estimated variances and covariances. To be able to accommodate the drastic changes in the magnitude of the estimates, the associated global shrinkage parameter $\kfQc_j$ can become too large to effectively shrink noise in calmer periods. In contrast, the dynamic triple gamma can adapt to the locally differing variance requirements of a given state. An example of this can be seen in Figure~\ref{fig_DPW_ADS}, which displays the marginal time-varying covariance of two returns, where the estimate from the static approach is much noisier than that under either dynamic specification. High resolution images of all variance-covariance matrices can be found \href{https://imgur.com/a/xoiGkFb}{here.}\footnote{In case the link is not clickable: https://imgur.com/a/xoiGkFb}


\begin{figure}[t!]
\vspace{-1em}
  \begin{center}
   \scalebox{0.7}
   {\includegraphics{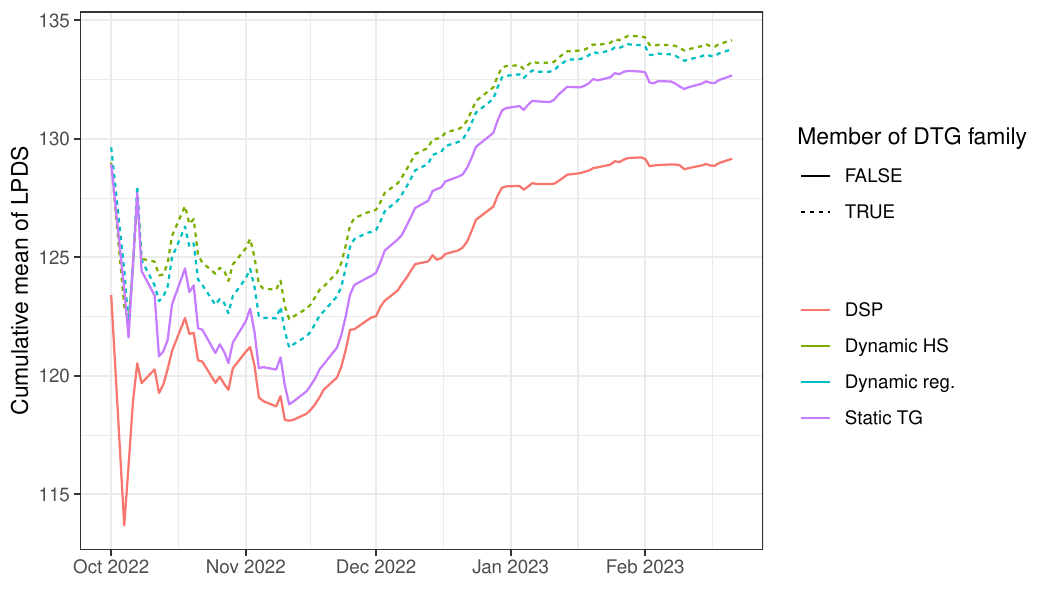}}
  \caption{\small Cumulative mean of one step ahead LPDS for the last 100 time points of the EURO STOXX data under static triple gamma (Static TG), dynamic triple gamma
  (Dynamic reg.), 
  dynamic horseshoe (Dynamic HS) and dynamic shrinkage process (DSP) of \cite{kow-etal:dyn}. \label{fig_cummean_LPDS}}
  \end{center}
  \vspace{-1em}
\end{figure}

\begin{figure}[t!]
\vspace{-1em}
  \begin{center}
   \scalebox{0.65}
   {\includegraphics{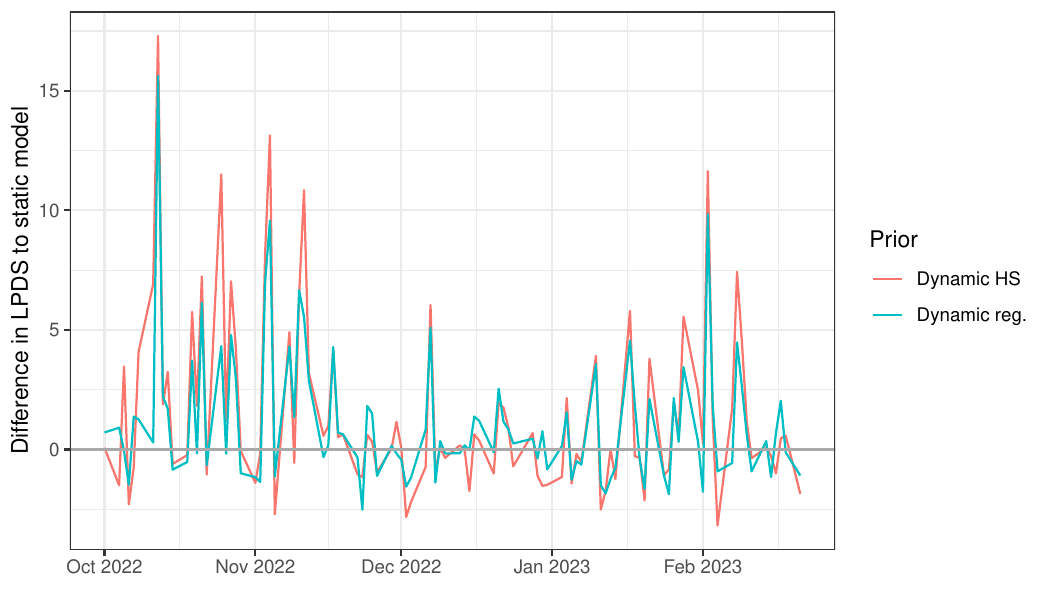}}
  \caption{\small Difference in one-step-ahead LPDS to static triple gamma for the last 100 time points of EURO STOXX data under DTG ($\cpsi_j=2.5$, Dynamic reg.) and dynamic horseshoe (Dynamic HS). \label{fig_diff_LPDS}}
  \end{center}
  \vspace{-1em}
\end{figure}

To benchmark the out-of-sample forecasting performance of our approach, we compare the one step ahead log-predictive density scores (LPDS) for the last 100 time points in the data set, as in \cite{bit-fru:ach}. The cumulative mean of this exercise can be found in Figure~\ref{fig_cummean_LPDS}. Both the slightly regularized dynamic triple gamma as well as the dynamic horseshoe outperform the static triple gamma, as is evidenced by the higher average LPDS throughout virtually the entire sample period. Furthermore, both the static and dynamic triple gamma priors clearly outperform the dynamic shrinkage process of \cite{kow-etal:dyn}. To gain intuition about how the dynamic triple gamma prior outperforms the static approach, it is instructive to look at Figure~\ref{fig_diff_LPDS}. It graphs the point-wise difference in LPDS vis-\`a-vis the static approach, for both the regularized dynamic triple gamma, as well as the dynamic horseshoe. It is interesting to note that the dynamic approaches forecasts better only in around $60 \%$ of time points. However, whenever the forecasting performance of the dynamic approach is worse than the static approach, it is usually not worse by much. In contrast, when the forecasting performance is better, it is often substantially so, leading to forecasting performance that is, on average, much better than under the static approach.

\vspace*{-5mm}

\section{Concluding remarks} \label{conc}

This paper introduced the dynamic triple gamma (DTG) prior as a novel shrinkage process for time-varying parameter (TVP) models. By generalizing the triple gamma prior to a stochastic process, the DTG approach balances local adaptivity and regularization, capturing time-varying innovation variances while remaining computationally tractable. The DTG prior’s theoretical properties were explored, clarifying connections to existing approaches like the NGAR and dynamic shrinkage processes. An application to the EURO STOXX 50 index demonstrated its utility, capturing complex patterns in time-varying (co)-variances and outperforming competing approaches in predictive performance. The DTG prior offers a flexible, interpretable, and computationally efficient approach to modeling time-varying parameters in Bayesian frameworks. Its combination of global shrinkage, local adaptivity, and time-varying dependence makes it a powerful tool for applications in econometrics, financial modeling, and beyond.

Future work could explore sharing information across equations for parameters $\rhotr_1, \dots, \rhotr_d$ via a hierarchical prior. For example, clustering these parameters with finite mixture models could encode prior beliefs about groups of covariates with similar dependence structures, enhancing shared learning.

\noindent {\bf Disclosure statement:} The authors report there are no competing interests to declare.

\vspace*{4mm}
\centerline{{\large\bf SUPPLEMENTARY MATERIAL}}
\noindent {\bf Appendix:} Appendix with proofs and additional technical details (PDF format).

\vspace*{-3mm}

\renewcommand{\baselinestretch}{1.1}

\bibliography{Manuscript_V2/sylvia_kyoto, Manuscript_V2/additional_bib}

\begin{thebibliography}{}

\bibitem[\protect\citeauthoryear{Bitto and Fr\"{u}hwirth-Schnatter}{Bitto and
  Fr\"{u}hwirth-Schnatter}{2019}]{bit-fru:ach}
Bitto, A. and S.~Fr\"{u}hwirth-Schnatter (2019).
\newblock Achieving shrinkage in a time-varying parameter model framework.
\newblock {\em Journal of Econometrics\/}~{\em 210\/}(1), 75--97.

\bibitem[\protect\citeauthoryear{Cadonna, Fr{\"u}hwirth-Schnatter, and
  Knaus}{Cadonna et~al.}{2020}]{cad-etal:tri}
Cadonna, A., S.~Fr{\"u}hwirth-Schnatter, and P.~Knaus (2020).
\newblock Triple the gamma -- {A} unifying shrinkage prior for variance and
  variable selection in sparse state space and {TVP} models.
\newblock {\em Econometrics\/}~{\em 8\/}(2), 20.

\bibitem[\protect\citeauthoryear{Carvalho, Polson, and Scott}{Carvalho
  et~al.}{2010}]{car-etal:hor}
Carvalho, C.~M., N.~G. Polson, and J.~G. Scott (2010).
\newblock The horseshoe estimator for sparse signals.
\newblock {\em Biometrika\/}~{\em 97\/}(2), 465--480.

\bibitem[\protect\citeauthoryear{Fr{\"u}hwirth-Schnatter}{Fr{\"u}hwirth-Schnatter}{2006}]{fru:book}
Fr{\"u}hwirth-Schnatter, S. (2006).
\newblock {\em Finite Mixture and Markov Switching Models}.
\newblock New York: Springer.

\bibitem[\protect\citeauthoryear{Fr{\"u}hwirth-Schnatter and
  Wagner}{Fr{\"u}hwirth-Schnatter and Wagner}{2010}]{fru-wag:sto}
Fr{\"u}hwirth-Schnatter, S. and H.~Wagner (2010).
\newblock Stochastic model specification search for {G}aussian and partially
  non-{G}aussian state space models.
\newblock {\em Journal of Econometrics\/}~{\em 154\/}(1), 85--100.

\bibitem[\protect\citeauthoryear{George and {McC}ulloch}{George and
  {McC}ulloch}{1993}]{geo-mcc:var}
George, E.~I. and R.~{McC}ulloch (1993).
\newblock Variable selection via {G}ibbs sampling.
\newblock {\em Journal of the American Statistical Association\/}~{\em
  88\/}(423), 881--889.

\bibitem[\protect\citeauthoryear{Gourieroux and Jasiak}{Gourieroux and
  Jasiak}{2006}]{gou-jas:aut}
Gourieroux, C. and J.~Jasiak (2006).
\newblock Autoregressive gamma processes.
\newblock {\em Journal of Forecasting\/}~{\em 25\/}(2), 129--152.

\bibitem[\protect\citeauthoryear{Griffin and Brown}{Griffin and
  Brown}{2010}]{gri-bro:inf}
Griffin, J.~E. and P.~J. Brown (2010).
\newblock Inference with normal-gamma prior distributions in regression
  problems.
\newblock {\em Bayesian Analysis\/}~{\em 5\/}(1), 171--188.

\bibitem[\protect\citeauthoryear{Griffin and Brown}{Griffin and
  Brown}{2017}]{gri-bro:hie}
Griffin, J.~E. and P.~J. Brown (2017).
\newblock Hierarchical shrinkage priors for regression models.
\newblock {\em Bayesian Analysis\/}~{\em 12\/}(1), 135--159.

\bibitem[\protect\citeauthoryear{Hauzenberger, Huber, Koop, and
  Mitchell}{Hauzenberger et~al.}{2022}]{hau-etal:tree}
Hauzenberger, N., F.~Huber, G.~Koop, and J.~Mitchell (2022).
\newblock Bayesian modeling of time-varying parameters using regression trees.
\newblock {\em arXiv preprint arXiv:2209.11970\/}.

\bibitem[\protect\citeauthoryear{Hauzenberger, Huber, Koop, and
  Onorante}{Hauzenberger et~al.}{2022}]{hau-etal:fas}
Hauzenberger, N., F.~Huber, G.~Koop, and L.~Onorante (2022).
\newblock Fast and flexible {B}ayesian inference in time-varying parameter
  regression models.
\newblock {\em Journal of Business \& Economic Statistics\/}~{\em 40\/}(4),
  1904--1918.

\bibitem[\protect\citeauthoryear{Hosszejni and Kastner}{Hosszejni and
  Kastner}{2021}]{hos-kas:mod}
Hosszejni, D. and G.~Kastner (2021).
\newblock Modeling univariate and multivariate stochastic volatility in {R}
  with stochvol and factorstochvol.
\newblock {\em Journal of Statistical Software\/}~{\em 100\/}(12), 1–34.

\bibitem[\protect\citeauthoryear{Huber, Kastner, and Feldkircher}{Huber
  et~al.}{2019}]{hub-etal:sho}
Huber, F., G.~Kastner, and M.~Feldkircher (2019).
\newblock Should {I} stay or should {I} go? {A} latent threhold approach to
  large-scale mixture innovation models.
\newblock {\em Journal of Applied Econometrics\/}~{\em 34\/}(5), 621--640.

\bibitem[\protect\citeauthoryear{Huber, Koop, and Onorante}{Huber
  et~al.}{2021}]{hub-etal:ind}
Huber, F., G.~Koop, and L.~Onorante (2021).
\newblock Inducing sparsity and shrinkage in time-varying parameter models.
\newblock {\em Journal of Business \& Economic Statistics\/}~{\em 39\/}(2),
  669--683.

\bibitem[\protect\citeauthoryear{Huber and Pfarrhofer}{Huber and
  Pfarrhofer}{2021}]{hub-pfa:dyn}
Huber, F. and M.~Pfarrhofer (2021).
\newblock Dynamic shrinkage in time-varying parameter stochastic volatility in
  mean models.
\newblock {\em Journal of Applied Econometrics\/}~{\em 36\/}(2), 262--270.

\bibitem[\protect\citeauthoryear{Irie}{Irie}{2019}]{irie2019bayesian}
Irie, K. (2019).
\newblock Bayesian dynamic fused lasso.
\newblock {\em arXiv preprint arXiv:1905.12275\/}.

\bibitem[\protect\citeauthoryear{Jacquier, Polson, and Rossi}{Jacquier
  et~al.}{1994}]{jac-etal:bay_ana}
Jacquier, E., N.~G. Polson, and P.~E. Rossi (1994).
\newblock Bayesian analysis of stochastic volatility models.
\newblock {\em Journal of Business {\rm \&} Economic Statistics\/}~{\em
  12\/}(4), 371--389.

\bibitem[\protect\citeauthoryear{Kalli and Griffin}{Kalli and
  Griffin}{2014}]{kal-gri:tim}
Kalli, M. and J.~E. Griffin (2014).
\newblock Time-varying sparsity in dynamic regression models.
\newblock {\em Journal of Econometrics\/}~{\em 178\/}(2), 779--793.

\bibitem[\protect\citeauthoryear{Kastner and Fr\"{u}hwirth-Schnatter}{Kastner
  and Fr\"{u}hwirth-Schnatter}{2014}]{kas-fru:anc}
Kastner, G. and S.~Fr\"{u}hwirth-Schnatter (2014).
\newblock Ancillarity-sufficiency interweaving strategy {(ASIS)} for boosting
  {MCMC} estimation of stochastic volatility models.
\newblock {\em Computational Statistics and Data Analysis\/}~{\em 76},
  408--423.

\bibitem[\protect\citeauthoryear{Knaus, Bitto-Nemling, Cadonna, and
  Fr{\"u}hwirth-Schnatter}{Knaus et~al.}{2021}]{knaus2021shrinkage}
Knaus, P., A.~Bitto-Nemling, A.~Cadonna, and S.~Fr{\"u}hwirth-Schnatter (2021).
\newblock {Shrinkage in the time-varying parameter model framework using the R
  package shrinkTVP}.
\newblock {\em Journal of Statistical Software\/}~{\em 100\/}(13).

\bibitem[\protect\citeauthoryear{Knaus, Bitto-Nemling, Cadonna, and
  Fr{\"{u}}hwirth-Schnatter}{Knaus et~al.}{2024}]{kna-etal-new:shr_TVP}
Knaus, P., A.~Bitto-Nemling, A.~Cadonna, and S.~Fr{\"{u}}hwirth-Schnatter
  (2024).
\newblock {\em shrinkTVP: Efficient Bayesian Inference for Time-Varying
  Parameter Models with Shrinkage}.
\newblock R package version 3.0.1.

\bibitem[\protect\citeauthoryear{Koop and Korobilis}{Koop and
  Korobilis}{2012}]{koo-kor:for}
Koop, G. and D.~Korobilis (2012).
\newblock Forecasting inflation using dynamic model averaging.
\newblock {\em International Economic Review\/}~{\em 53\/}(3), 867--886.

\bibitem[\protect\citeauthoryear{Kowal, Matteson, and Ruppert}{Kowal
  et~al.}{2019}]{kow-etal:dyn}
Kowal, D.~R., D.~S. Matteson, and D.~Ruppert (2019).
\newblock Dynamic shrinkage processes.
\newblock {\em Journal of the Royal Statistical Society, Ser. B\/}~{\em
  81\/}(4), 781--804.

\bibitem[\protect\citeauthoryear{Kraus}{Kraus}{1956}]{kraus1956graphs}
Kraus, E. (1956).
\newblock Graphs of cumulative residuals.
\newblock {\em Quarterly Journal of the Royal Meteorological Society\/}~{\em
  82\/}(351), 96--98.

\bibitem[\protect\citeauthoryear{Lopes, McCulloch, and Tsay}{Lopes
  et~al.}{2022}]{lop-etal:par}
Lopes, H.~F., R.~E. McCulloch, and R.~S. Tsay (2022).
\newblock Parsimony inducing priors for large scale state–space models.
\newblock {\em Journal of Econometrics\/}~{\em 230\/}(1), 39--61.

\bibitem[\protect\citeauthoryear{Makalic and Schmidt}{Makalic and
  Schmidt}{2016}]{mak-sch:sim}
Makalic, E. and D.~F. Schmidt (2016).
\newblock A simple sampler for the horseshoe estimator.
\newblock {\em {IEEE} Signal Processing Letters\/}~{\em 23\/}(1), 179--182.

\bibitem[\protect\citeauthoryear{{McC}ausland, Miller, and
  Pelletier}{{McC}ausland et~al.}{2011}]{mcc-etal:sim}
{McC}ausland, W.~J., S.~Miller, and D.~Pelletier (2011).
\newblock Simulation smoothing for state space models: {A} computational
  efficiency analysis.
\newblock {\em Computational Statistics and Data Analysis\/}~{\em 55\/}(1),
  199--212.

\bibitem[\protect\citeauthoryear{Nakajima and West}{Nakajima and
  West}{2013}]{nak-wes:bay_ana}
Nakajima, J. and M.~West (2013).
\newblock Bayesian analysis of latent threshold dynamic models.
\newblock {\em Journal of Business {\rm \&} Economic Statistics\/}~{\em
  31\/}(2), 151--164.

\bibitem[\protect\citeauthoryear{Petris, Petrone, and Campagnoli}{Petris
  et~al.}{2009}]{pet-etal:dyn}
Petris, G., S.~Petrone, and P.~Campagnoli (2009).
\newblock {\em Dynamic Linear Models with R}.
\newblock New York: Springer.

\bibitem[\protect\citeauthoryear{Ro\v{c}kov\'{a} and McAlinn}{Ro\v{c}kov\'{a}
  and McAlinn}{2021}]{roc-mca:dyn}
Ro\v{c}kov\'{a}, V. and K.~McAlinn (2021).
\newblock Dynamic variable selection with spike-and-slab process priors.
\newblock {\em Bayesian Analysis\/}~{\em 16\/}(1), 233--269.

\bibitem[\protect\citeauthoryear{Uribe and Lopes}{Uribe and
  Lopes}{2020}]{uri-hed:dyn}
Uribe, P.~W. and H.~F. Lopes (2020).
\newblock Dynamic sparsity on dynamic regression models.
\newblock {\em arXiv preprint arXiv:2009.14131\/}.

\bibitem[\protect\citeauthoryear{{van D}yk and Park}{{van D}yk and
  Park}{2008}]{van-par:par}
{van D}yk, D.~A. and T.~Park (2008).
\newblock Partially collapsed {G}ibbs samplers: {T}heory and methods.
\newblock {\em Journal of the American Statistical Association\/}~{\em
  103\/}(482), 790--796.

\end{thebibliography}

\newpage

\begin{appendices}
\setcounter{page}{1}
\renewcommand{\baselinestretch}{1.1}

\begin{center}
{\bf \Large Appendix for "The Dynamic Triple Gamma Prior as a Shrinkage Process Prior for Time-Varying Parameter Models"}\\[1cm]
\end{center}

\section{Details on local level model} \label{sec_local_level}

The local level model used as an example in Section~\ref{sec_intro} is specified as:
\begin{eqnarray*}
     & & y_{t} =  \beta_{t} +  \error_{t}, \qquad
\error_{t} \sim \Normal{0, \sigmaerr},  \\
&&  \beta_{t} = \beta_{t-1} + w_{t}, \qquad   w_{t}  \sim \Normal{0, \kfQc \psis_t},
\end{eqnarray*}
where $\beta_0 \sim \Normal{\beta, \kfQc \psis_0}$. In the static case, $\psis_t = 1$, with $\kfQc$ assigned a triple gamma prior ($a^\xi = c^\xi = 0.5$) and $\beta^2$ also assigned a triple gamma prior ($a^\tau = c^\tau = 0.5$). The prior on $\sigma^2$ is conditionally inverse gamma: $\sigma^2 | C_0 \sim \Gammainv{c_0, C_0}$, where $C_0 \sim \Gammad{g_0, G_0}$, with $c_0 = 2.5$, $g_0 = 5$, and $G_0 = g_0 / (c_0 - 1)$.

In the dynamic case, $\kfQc$, $\beta$ and $\sigma^2$ retain the same priors, but $\psis_t$ is placed under the process prior defined in \eqref{innov}-\eqref{kappa} with $a = c = 0.5$. Additionally, $\rhotr_j$ is assigned a GB1 prior as defined in \eqref{rho_prior}, with hyperparameters $a_\rho = 1$, $b_\rho = 0.95$, $\alpha_\rho = 0.5$, and $\beta_\rho = 0.5$.

\section{Details on non-standard distributions} \label{nonstand_dist}

\subsection{The beta-prime distribution} 

If $ X \sim \Betapr{a ,c}$ follows a beta-prime distribution, then it has  pdf $$ f(x; a, c)= \frac{1}{\Betafun{a,c}}  \frac{x ^{a-1}}{(1+x)^{a+c}}.$$
 The r.v.  $Y= X/(1+X)$ follows the $ \Betadis{a ,c}$-distribution, meaning that the odds  $ X  =Y/(1-Y)$ of a $ \Betadis{a ,c}$-distribution follow the $  \Betapr{a ,c}$-distribution. Furthermore, $Y= \frac{c}{a} X$  follows the $ \Fd{2a ,2c}$-distribution. Hence, the beta prime distribution can be represented as the ratio of two independent gamma distributions:
  \begin{eqnarray*} 
  \Betapr{a ,c} = _d \frac{\Gammad{a,1}}{\Gammad{c,1}}.
   \end{eqnarray*}
   This follows immediately from
    \begin{eqnarray*} 
    \Betapr{a ,c} = 
    \frac{a}{c} \Fd{2 \atau, 2 \ctau} =_d  \frac{a}{c} \frac{\Gammad{\atau,1}/\atau}{\Gammad{\ctau,1}/\ctau} = \frac{\Gammad{a,1}}{\Gammad{c,1}}.
 \end{eqnarray*}

\subsection{The generalized beta-prime distribution}

If $X \sim \GenBetapr{\alpha, \beta, p, q}$, then it has pdf
  \begin{eqnarray*} 
f(x ; \alpha, \beta, p, q)=\frac{p\left(\frac{x}{q}\right)^{\alpha p-1}\left(1+\left(\frac{x}{q}\right)^p\right)^{-\alpha-\beta}}{q B(\alpha, \beta)}.
 \end{eqnarray*}

 \subsection{The Z-distribution} 
If $X\sim Z(\alpha, \beta, \mu_z, \sigma_z)$ follows a Z-distribution, then it has pdf
\begin{equation*} 
f(x; \alpha, \beta, \mu_z, \sigma_z)  = \frac{\exp\left(\frac{z - \mu_z}{\sigma_z}\right)^\alpha\left(1 + \exp\left(\frac{z - \mu_z}{\sigma_z}\right)\right)^{-(\alpha + \beta)}}{\sigma_z \Betafun{\alpha, \beta}}.
\end{equation*}

\section{Further transition densities of the dynamic triple gamma} \label{state_spaces}

\subsection{The transition density \texorpdfstring{$p(\lambdac_{jt} |\lambdac_{j,t-1}, \apsi_j , \cpsi_j, \rhotr_j)$}{p(lambda jt|lambda j,t-1, a, c, rho)}} \label{tranlamlam}

The transition density $\lambdac_{jt} |\lambdac_{j,t-1}, \apsi_j , \cpsi_j, \rhotr _j$ is marginalized
 w.r.t. $\kappaPG_{jt}|\lambdac_{j,t-1}, \apsi_j, \cpsi_j ,\rhotr _j$.
 The transition density $\lambdac_{jt} |\lambdac_{j,t-1}, \apsi_j , \cpsi_j, \rhotr _j$ arises from the rescaled noncentral $\Chinon{2\apsi_j}{\Chinoncen_{jt}}$-distribution with $2\apsi_j$
      degrees of freedom and non-centrality parameter
      \begin{eqnarray*} 
       \Chinoncen_{jt}= \frac{2\apsi_j}{\cpsi_j} \frac{\rhotr _j}{1- \rhotr _j} \lambdac_{j,t-1}.
      \end{eqnarray*}
      More specifically,
\begin{eqnarray} \label{prognon}
 p(\lambdac_{jt} |\lambdac_{j,t-1}, \apsi_j , \cpsi_j, \rhotr _j) =
 \frac{2\apsi_j}{\cpsi_j(1- \rhotr_j)}  f_{\Chinon{2\apsi_j}{\Chinoncen_{jt}}}
 \left( \lambdac_{jt} \frac{2\apsi_j}{\cpsi_j(1- \rhotr_j)} \right),
 \end{eqnarray}
 where $f_{\Chinon{2\apsi_j}{\Chinoncen_{jt}}}$ is the density of the noncentral
 $\Chinon{2\apsi_j}{\Chinoncen_{jt}}$-distribution.\footnote{The noncentral $\Chi$ distribution, $\rvY \sim
\Chinon{\nu}{\Chinoncen}$ can be seen as Poisson-weighted mixture of $\Chi$ distributions:
\begin{eqnarray} \label{densnchi}
 f_{\Chinon{\nu}{\Chinoncen}} (\rvYy)  =
 \sum_{k=0}^\infty \frac{\e ^{-\Chinoncen/2} (\Chinoncen/2) ^k}{k!} f_{\Chisqu{\nu + 2k}} (\rvYy),
\end{eqnarray} where $f_{\Chisqu{\nu + 2k}} (\rvYy)$ is the density of a $\Chisqu{\nu + 2k}$-distribution.}
       Note that $\Chinoncen_{jt}$ is twice the intensity of the Poisson distribution
      $\kappaPG_{jt}|\lambdac_{j,t-1}, \apsi_j, \cpsi_j ,\rhotr _j $ \SFS{given in \eqref{kappa}}, while
      $\frac{2\apsi_j}{\cpsi_j(1- \rhotr_j)}$ is twice the  scale parameter of
      the gamma distribution $\lambdac_{jt} |\kappaPG_{jt}, \apsi_j, \cpsi_j, \rhotr _j$ \SFS{given in \eqref{lambda}}.
      The proof of (\ref{prognon}) is based on rewriting the gamma density of the r.v. $\lambdac_{jt} |\kappaPG_{jt}, \apsi_j, \cpsi_j, \rhotr _j$ as
      \begin{eqnarray*}
       \lambdac_{jt}  =  \frac{\cpsi_j(1- \rhotr_j)}{2\apsi_j} \rvX_{jt} , \qquad  \rvX _{jt}| \kappaPG_{jt}=k, \apsi_j  \sim \Gammad{2(\apsi_j +k)/2,1/2} \sim \Chisqu{2\apsi_j + 2k}.
      \end{eqnarray*}
      Since $\kappaPG_{jt}|\lambdac_{j,t-1}, \apsi_j, \cpsi_j ,\rhotr _j \sim \Poisson{\frac{\apsi_j\rhotr _j}{\cpsi_j (1- \rhotr _j)}\lambdac_{j,t-1} } =
      \Poisson{\Chinoncen_{jt}/2}$, it follows immediately that
       $$ \rvX _{jt}|\lambdac_{j,t-1}, \apsi_j, \cpsi_j ,\rhotr _j \sim  \Chinon{2\apsi_j}{\Chinoncen_{jt}}$$ follows
       a noncentral $\Chi$-distribution, see also (\ref{densnchi}).
      (\ref{prognon})  follows immediately from the law of transformation of densities
      applied to $\rvX_{jt}= \frac{2\apsi_j}{\cpsi_j(1- \rhotr_j)}  \lambdac_{jt}$.

\subsection{The transition density \texorpdfstring{$p(\kappaPG_{jt} | \kappaPG_{j,t-1}, \apsi_j , \cpsi_j, \rhotr _j)$}{p(kappa jt|kappa j,t-1, a, c, rho)}}
 
   \cite{gou-jas:aut} show that following marginal transition density $\kappaPG_{jt} | \kappaPG_{j,t-1}, \apsi_j, \cpsi_j, \rhotr _j  $ holds for $t=2, \ldots, T$:
 \begin{eqnarray} \label{transKj}
\kappaPG_{jt}  | \apsi_j, \rhotr_j, \kappaPG_{j,t-1} \sim \NegBin{\apsi_j + \kappaPG_{j,t-1}, \piNB _j}, \quad \piNB_j= \frac{1 }{1+\rhotr_j },
 \end{eqnarray}
 while the initial distribution at $t=1$ is:
  \begin{eqnarray} \label{transKj0}
\kappaPG_{j1} | \apsi_j, \rhotr_j \sim \NegBin{\apsi_j , 1-\rhotr_j } .
 \end{eqnarray}
This can also be seen directly, as when $\rvX |\rvY \sim  \Poisson{\thetatr \rvY}$
and $\rvY  \sim \Gammad{\alpha ,\beta}$, then $\rvX \sim \NegBin{\alpha, \piNB}$, where
$ \piNB=\frac{\beta}{\beta + \thetatr}, $
 i.e.
 \begin{eqnarray} \label{negdens}
 \Prob{\rvX =\rvXx}=\frac{\Gamfun{\alpha + \rvXx}}{ \rvXx ! \, \Gamfun{\alpha}} \piNB ^ \alpha (1-\piNB)^{\rvXx } .
 \end{eqnarray} 

 \subsection{The transition density \texorpdfstring{$p(\kappaPG_{jt} | \kappaPG_{j,t-1}, \psis_{j,t-1},\apsi_j , \cpsi_j, \rhotr _j)$}{p(kappa jt|kappa j,t-1, psi j,t-1, a, c, rho)}}
 
The transition density (\ref{transKj}) is obtained by marginalizing over $\lambdac_{j,t-1}$. When doing so, one introduces dependence between $\kappaPG_{jt}$ and $\psis_{j,t-1}$ (see also Figure~\ref{tg_graph_alt}). The  transition density 
   $\kappaPG_{jt}  | \kappaPG_{j,t-1},  \psis_{j,t-1} , \rhotr_j, \apsi_j, \cpsi_j$ is defined as:
   \begin{eqnarray*} \label{KjtpostB}
 p(\kappaPG_{jt}  | \kappaPG_{j,t-1},\psis_{j,t-1}, \apsi_j, \cpsi_j, \rhotr_j) =
 \int   p(\kappaPG_{jt}  |  \lambdac_{j,t-1}, \apsi_j, \cpsi_j, \rhotr_j) p( \lambdac_{j,t-1}| \kappaPG_{j,t-1},\psis_{j,t-1}, \apsi_j, \cpsi_j, \rhotr_j) \,
 d \, \lambdac_{j,t-1}.
    \end{eqnarray*}
Using\footnote{Same derivation as for $\lambdac_{jT}|\kappaPG_{jT},\psis_{jT}, \apsi_j, \cpsi_j, \rhotr_j $  in full conditional Gibbs, see Appendix~\ref{FCG}}
\begin{eqnarray} \label{KjtpostC}
\lambdac_{j,t-1}| \kappaPG_{j,t-1},\psis_{j,t-1}, \apsi_j, \cpsi_j, \rhotr_j \sim
\Gammad{\apsi_j + \cpsi_j + \kappaPG_{j,t-1},
\frac{\apsi_j}{\cpsi_j}\frac{1}{1- \rhotr_j } + \frac{1}{\psis_{j,t-1} }},
\end{eqnarray}
in combination with (\ref{negdens}), we obtain following negative binomial distribution:
\begin{eqnarray} \label{KjtpostD}
\kappaPG_{jt}  | \kappaPG_{j,t-1},  \psis_{j,t-1} , \rhotr_j, \apsi_j, \cpsi_j &\sim &
\NegBin{\apsi_j + \cpsi_j + \kappaPG_{j,t-1}, \piNB _{j,t-1}}, \\  \nonumber
 \piNB_{j,t-1} & = &
 \frac{\apsi_j\psis_{j,t-1} + \cpsi_j (1- \rhotr_j) }{(1+\rhotr_j) \apsi_j\psis_{j,t-1} + \cpsi_j (1- \rhotr_j) }.
\end{eqnarray}
In comparison to the transition density (\ref{transKj}), we find that $\cpsi_j$ adds to the \lq\lq number of trials\rq\rq\ (the first parameter), while the \lq\lq success probability \rq\rq\ $\piNB_{j,t-1}$ (the second parameter) is time-varying and depends on the past scale $\psis_{j,t-1}$, as opposed to the fixed
success probability $\piNB_{j}$ in (\ref{transKj}). 
For  $\rhotr_j=1$ and $\psis_{j,t-1}=\infty$,
the two probabilities coincide,
$\piNB_{j,t-1}=\piNB_{j}$,  otherwise $\psis_{j,t-1}$ will impact $\piNB_{j,t-1}$.

\subsection{The transition density \texorpdfstring{$p(\psis_{jt} | \kappaPG_{jt}, \apsi_j , \cpsi_j,\rhotr _j)$}{p(psi jt|kappa jt, a, c, rho)}}
       \label{transpsikap}

Combining representation \eqref{rephor} with the conditional density $\lambdac_{jt} |\kappaPG_{jt}, \apsi_j, \cpsi_j, \rhotr _j \sim \Gammad{\apsi_j  + \kappaPG_{jt} , \frac{\apsi_j}{\cpsi_j}\frac{1}{1- \rhotr_j }}$, it is possible to
 derive the following closed form representation of $p( \psis_{jt} | \kappaPG_{jt}, \apsi_j, \cpsi_j, \rhotr _j )$:
  \begin{eqnarray} \label{poshflde}
  && p( \psis_{jt} | \kappaPG_{jt}, \apsi_j, \cpsi_j, \rhotr _j ) = \\
&&\nonumber  \frac{1}{\Betafun{\apsi_j+  \kappaPG_{jt}, \cpsi_j}}  \frac{\apsi_j}{\cpsi_j (1- \rhotr_j )}
\left( \frac{\apsi_j\psis_{jt}}{\cpsi_j (1- \rhotr_j )}  \right) ^{ \apsi_j+  \kappaPG_{jt}-1}
\left( \frac{\apsi_j\psis_{jt}}{\cpsi_j (1- \rhotr_j )} +1 \right) ^{ -({\apsi_j + \cpsi_j+  \kappaPG_{jt}})}.
\end{eqnarray}
This implies that  $\psis_{jt} | \kappaPG_{jt}, \apsi_j, \cpsi_j, \rhotr _j \sim 
\GenBetapr{\apsi_j + \kappaPG_{jt}, \cpsi_j, 1, \frac{\cpsi_j(1 - \rhotr_j )}{\apsi_j}}$, where 
$\GenBetapr{\alpha, \beta, p, q}$
is the generalized beta prime distribution.

\SFS{The proof of (\ref{poshflde}) is striaghtforward:}
 \begin{eqnarray*}
&&p(\psis_{jt} | \kappaPG_{jt},  \apsi_j, \cpsi_j, \rhotr _j) =
\int   p(\psis_{jt} | \kappaPG_{jt}, \lambdac_{jt} , \apsi_j, \cpsi_j, \rhotr _j ) p(\lambdac_{jt} |\kappaPG_{jt}, \apsi_j, \cpsi_j, \rhotr _j) \, d \, \lambdac_{jt} = \\
&&  \frac{1}{\Gamfun{\cpsi_j} \Gamfun{\apsi_j+  \kappaPG_{jt}}} \left(\frac{1}{\psis_{jt}} \right) ^{ \cpsi_j+1}
\left( \frac{\apsi_j}{\cpsi_j (1- \rhotr_j )} \right) ^{ \apsi_j+  \kappaPG_{jt}} \cdot \\
& & \qquad \cdot
\int   \lambdac_{jt} ^{\cpsi_j} \exp \left( - \frac{\lambdac_{jt}}{\psis_{jt}}  \right) \lambdac_{jt} ^{ \apsi_j+  \kappaPG_{jt}-1}  \exp \left( - \frac{\lambdac_{jt}\apsi_j}{\cpsi_j (1- \rhotr_j )} \right)    \, d \, \lambdac_{jt} = \\
&& \frac{\Gamfun{\apsi_j + \cpsi_j+  \kappaPG_{jt}}}{\Gamfun{\cpsi_j} \Gamfun{\apsi_j+  \kappaPG_{jt}}} \left(\frac{1}{\psis_{jt}} \right) ^{ \cpsi_j+1}
\left( \frac{\apsi_j}{\cpsi_j (1- \rhotr_j )}  \right) ^{ \apsi_j+  \kappaPG_{jt}}
\left( \frac{\apsi_j}{\cpsi_j (1- \rhotr_j )}  + \frac{1}{\psis_{jt}}  \right) ^{ -({\apsi_j + \cpsi_j+  \kappaPG_{jt}})} = \\
&&  \frac{1}{\Betafun{\apsi_j+  \kappaPG_{jt}, \cpsi_j}}  \frac{\apsi_j}{\cpsi_j (1- \rhotr_j )}
\left( \frac{\apsi_j\psis_{jt}}{\cpsi_j (1- \rhotr_j )}  \right) ^{ \apsi_j+  \kappaPG_{jt}-1}
\left( \frac{\apsi_j\psis_{jt}}{\cpsi_j (1- \rhotr_j )} +1 \right) ^{ -({\apsi_j + \cpsi_j+  \kappaPG_{jt}})}.
\end{eqnarray*}

\eqref{poshflde} together with \eqref{KjtpostD} and \eqref{transKj0} defines a new state space model with the same properties as the one introduced in Section~\ref{sec_dtg}, albeit with less convenient conditional densities involved. A representation of this new state space model as a directed graph can be found in Figure~\ref{tg_graph_alt}.

\subsection{The transition density \texorpdfstring{$p(\psis_{jt}|\lambdac_{j,t-1}, \apsi_j , \cpsi_j,\rhotr _j)$}{p(psi jt|lambda j,t-1, a, c, rho)}} \label{transpsilam}

  The transition density  $\psis_{jt}|\lambdac_{j,t-1}, \apsi_j , \cpsi_j,\rhotr _j $  is marginalized
 w.r.t. $\kappaPG_{jt}|\lambdac_{j,t-1}, \apsi_j, \cpsi_j ,\rhotr _j$.
The transition density  $\psis_{jt}|\lambdac_{j,t-1}, \apsi_j , \cpsi_j,\rhotr _j $ arises from the rescaled noncentral
      $\Fdnon{2\apsi_j,2\cpsi_j}{\Chinoncen_{jt}}$-distribution with 
       $2\apsi_j$ and $2\cpsi_j$
      degrees of freedom respectively, and the same non-centrality parameter $\Chinoncen_{jt}$ defined
      in (\ref{prognon}).
         More specifically,
\begin{eqnarray} \label{progFdno}
p(\psis_{jt}|{\lambdac_{j,t-1}}, \apsi_j , \cpsi_j,\rhotr _j) =
 \frac{1}{1- \rhotr_j}  f_{\Fdnon{2\apsi_j,2\cpsi_j}{\Chinoncen_{jt}}}
 \left(\frac{\psis_{jt}}{1- \rhotr_j} \right),
 \end{eqnarray}
 where $f_{\Fdnon{2\apsi_j,2\cpsi_j}{\Chinoncen_{jt}}}$ is the density of the noncentral
 $\Fdnon{2\apsi_j,2\cpsi_j}{\Chinoncen_{jt}}$-distribution.\footnote{The noncentral F-distribution, $\rvY \sim
       \Fdnon{\nu_1,\nu_2}{\Chinoncen}$, has the density function
      \begin{eqnarray} \label{densFdnon}
 f_{\Fdnon{\nu_1,\nu_2}{\Chinoncen}} (\rvYy)  = \sum_{k=0}^{\infty} \frac{e^{-\Chinoncen / 2}(\Chinoncen / 2)^k}{B\left(\frac{\nu_2}{2}, \frac{\nu_1}{2}+k\right) k !}\left(\frac{\nu_1}{\nu_2}\right)^{\frac{\nu_1}{2}+k}\left(\frac{\nu_2}{\nu_2+\nu_1 f}\right)^{\frac{\nu_1+\nu_2}{2}+k} y^{\nu_1 / 2-1+k}.
\end{eqnarray}}

The proof of (\ref{progFdno}) is based on rewriting the density of the r.v. $\psis_{jt}|\kappaPG_{jt}, \apsi_j , \cpsi_j,\rhotr _j$ as
      \begin{eqnarray*}
       \psis_{jt}  =  (1- \rhotr_j) \rvX_{jt} , \qquad  \rvX _{jt}| \kappaPG_{jt}=k, \apsi_j, \cpsi_j  \sim \Fd{2(\apsi_j +k),2\cpsi_j}.
      \end{eqnarray*}
      This representations follows immediately from  the density of an F-distribution and (\ref{poshflde}).
      Since $\kappaPG_{jt}|\lambdac_{j,t-1}, \apsi_j, \cpsi_j ,\rhotr _j \sim \Poisson{\Chinoncen_{jt}/2}$, it follows immediately that
       $$ \rvX _{jt}|\lambdac_{j,t-1}, \apsi_j, \cpsi_j ,\rhotr _j \sim  \Fdnon{2\apsi_j,2\cpsi_j}{\Chinoncen_{jt}}$$ follows
       a noncentral F-distribution, see also (\ref{densFdnon}).
      (\ref{progFdno})  follows immediately from the law of transformation of densities
      applied to $\rvX_{jt}= \frac{\psis_{jt}}{1- \rhotr_j}$.

\section{Proofs} \label{proofs}

{\em Proof of Theorem~\ref{theo_cond}:} The proof is based on combining \eqref{progFdno} with the density $p(\lambdac_{jt}|\psis_{jt}, \apsi_j, \cpsi_j)$, the latter of which can be easily derived under the assumption that $\lambdac_{jt} \sim \Gammad{\apsi_j, \frac{\apsi_j}{\cpsi_j}}$ via Bayes theorem. To this end, note that
\begin{align*}
    p(\lambdac_{jt}|\psis_{jt}, \apsi_j, \cpsi_j) &\propto p(\psis_{jt}|\lambdac_{jt}, \apsi_j, \cpsi_j)p(\lambdac_{jt}|\apsi_j, \cpsi_j) \\
    &\propto \lambdac_{jt}^{\cpsi_j} e^{-\lambdac_{jt}\frac{1}{\psis_{jt}}} \lambdac_{jt}^{\apsi_j - 1}e^{-\lambdac_{jt}\frac{\apsi_j}{\cpsi_{jt}}} \\
    &=  \lambdac_{jt}^{\apsi_j + \cpsi_j - 1} e^{-\lambdac_{jt}\left(\frac{1}{\psis_{jt}} + \frac{\apsi_j}{\cpsi_j}\right)},
\end{align*}
from which it follows that
\begin{equation} \label{lampsi}
\lambdac_{jt}|\psis_{jt}, \apsi_j, \cpsi_j \sim \Gammad{\apsi_j + \cpsi_j, \frac{1}{\psis_{jt}} + \frac{\apsi_j}{\cpsi_j}}. 
\end{equation}
Taking this, combining it with \eqref{progFdno} and letting 
$$A = \frac{1}{1 - \rhotr_j}\frac{\left(\frac{1}{\psis_{j,t-1}} + \frac{\apsi_j}{\cpsi_j}\right)^{\apsi_j + \cpsi_j}}{\Gamfun{\apsi_j + \cpsi_j}}$$
and
$$C(k) = \frac{\left(\frac{\apsi_j\rhotr_j}{\cpsi_j(1 - \rhotr_j)}\right)^k\left(\frac{\apsi_j}{\cpsi_j}\right)^{\apsi_j + k}\left(\frac{\cpsi_j}{\cpsi_j + \frac{\apsi_j\psis_{jt}}{1 - \rhotr_j}}\right)^{\apsi_j + \cpsi_j + k}\left(\frac{\psis_{jt}}{1 - \rhotr_j}\right)^{\apsi_j + k - 1}}{\Betafun{\apsi_j + k, \cpsi_j}}$$
yields
\begin{align*}
    p(\psis_{jt}&|\psis_{j,t-1}, \apsi_j, \cpsi_j, \rhotr_j) = \int_{0}^\infty p(\psis_{jt}|\lambdac_{j,t-1}, \apsi_j, \cpsi_j, \rhotr_j)p(\lambdac_{j,t-1}|\psis_{j,t-1}, \apsi_j, \cpsi_j)d\lambdac_{j,t-1} \\
    &= A\int_{0}^\infty\left[\sum_{k = 0}^\infty C(k) e^{-\frac{\apsi_j}{\cpsi_j} \frac{\rhotr_j}{1-\rhotr_j}\lambdac_{j,t-1}}\lambdac_{j,t-1}^k\right]\lambdac_{j,t-1}^{\apsi_j + \cpsi_j - 1} e^{-\lambdac_{j,t-1}\left(\frac{1}{\psis_{j,t-1}} + \frac{\apsi_j}{\cpsi_j}\right)} d\lambdac_{j,t-1} \\
    &= A\sum_{k = 0}^\infty C(k) \int_{0}^\infty \lambdac_{j,t-1}^{\apsi_j + \cpsi_j + k -1} e^{-\left(\frac{\apsi_j}{\cpsi_j} \frac{1}{1-\rhotr_j} + \frac{1}{\psis_{j,t-1}}\right)\lambdac_{j,t-1}} d\lambdac_{j, t-1}\\
    &= A\sum_{k = 0}^\infty C(k) \frac{\Gamfun{\apsi_j + \cpsi_j + k}}{\left(\frac{\apsi_j}{\cpsi_j}\frac{1}{1 - \rhotr_j} + \frac{1}{\psis_{j,t-1}}\right)^{\apsi_j + \cpsi_j + k}}.
\end{align*}
Combined with the definition of the hypergeometric function and after some algebraic manipulations the above results in \eqref{psi_cond}.

{\em Proof of Theorem~\ref{expect_cond}:} The derivation of this result relies on the tower property combined with \eqref{progFdno} and $p(\lambdac_{j,t-1}|\psis_{j,t-1}, \apsi_j, \cpsi_j)$. First, note that 
\begin{align*}
    \mathbb E \left[ \psis_{jt} | \psis_{j,t-1}, \apsi_j, \cpsi_j, \rhotr_j \right] = \mathbb E \left[ \mathbb E \left[\psis_{jt}|\lambdac_{j,t-1}, \apsi_j, \cpsi_j \right] | \psis_{j,t-1}, \apsi_j, \cpsi_j, \rhotr_j \right].
\end{align*}
Based on \eqref{progFdno} and the properties of the noncentral F-distribution (which requires that $\cpsi_j > 1$ for the expected value to exist), it can be deduced that 
\begin{align*}
    \mathbb E &\left[\psis_{jt} |\lambdac_{j,t-1}, \apsi_j, \cpsi_j \right] = (1 - \rhotr_j)  \mathbb E \left[\frac{\psis_{jt}}{(1 - \rhotr_j)}|\lambdac_{j,t-1}, \apsi_j, \cpsi_j \right] \\
    &= (1 - \rhotr_j) \frac{\cpsi_j}{\cpsi_j - 1} + \rhotr_j \frac{1}{\cpsi_j - 1}\mathbb E \left[\lambdac_{j,t-1} |\psis_{j,t-1}, \apsi_j, \cpsi_j \right],
\end{align*}
which, through the use of the properties of the gamma distribution leads to 
\begin{align*}
    \mathbb E &\left[\psis_{jt} |\lambdac_{j,t-1}, \apsi_j, \cpsi_j \right] = (1 - \rhotr_j) \frac{\cpsi_j}{\cpsi_j - 1} + \rhotr_j \frac{1}{\cpsi_j - 1}\frac{\apsi_j + \cpsi_j}{\frac{1}{\psis_{j,t-1}} + \frac{\apsi_j}{\cpsi_j}}.
\end{align*}
Following some algebraic manipulations, this results in \eqref{expec_cond}.


\section{More about prior choices}  \label{sec_prior_details}

Throughout the work, we place triple gamma priors on both $\betaci{j}^2 \sim TG(a^\tau, c^\tau, \lambda_B^2)$ and  $\kfQc_j \sim TG(a^\xi, c^\xi, \kappaBj{B}^2)$. As $\kfQc_j$ acts as the global shrinkage parameter and $\betaci{j}$ is the mean of the initial value of state $\{\betaci{jt}\}$, this prior setup allows the model to effectively differentiate between time-varying, static and excluded covariates. Further, we place scaled beta priors on $a^\tau, a^\xi, c^\tau$ and $c^\xi$:
\begin{equation} \label{hyperprior_ac}
	2 a^\xi \sim \mathcal B (\alpha_{a^\xi},  \beta_{a^\xi}) ,
	\quad 2 c^\xi \sim \mathcal B (\alpha_{c^\xi},  \beta_{c^\xi}), \qquad 2 a^\tau \sim \mathcal B (\alpha_{a^\tau},  \beta_{a^\tau}) ,
	\quad 2 c^\tau \sim \mathcal B (\alpha_{c^\tau},  \beta_{c^\tau}).
\end{equation}
This ensures that all shape parameters remain in the range $(0, 0.5)$, thereby preserving the strong pole around the origin as well as the heavy tails for the global shrinkage parameter $\kfQc_j$. 

We further follow the arguments in \cite{cad-etal:tri} and place scaled F priors on the global shrinkage parameters $\lambda_B^2$ and $\kappaBj{B}^2$. Specifically,
\begin{equation} \label{hyperprior_kappa}
    \frac{\kappaBj{B}^2}{2}\mid a^\xi, c^\xi \sim \Fd{2a^\xi, 2c^\xi}, \quad \frac{\lambda_{B}^2}{2}\mid a^\tau, c^\tau \sim \Fd{2a^\tau, 2c^\tau}.
\end{equation}
Finally, a prior has to be specified for the 
\SFSnew{error variance $\sigmaerr_t$ in}
the observation equation. Here, two cases have to be differentiated. In the homoscedastic $\sigmaerr_t = \sigmaerr$ case, we place a hierarchical prior on $\sigma^2$ , where the scale of an inverse gamma follows a gamma distribution:
\begin{eqnarray} \label{eq:priorsigma}
	\sigma^2|C_0 \sim \Gammainv{c_0,C_0}, \qquad  C_0 \sim \Gammad{g_0,G_0},
\end{eqnarray}
with hyperparameters $c_0$, $g_0$, and $G_0$.
In the case of stochastic volatility, we follow \citet{kas-fru:anc} in the choice of priors on the parameters $\mu$, $\phi$ and $\sigma^2_\eta$ in Equation~\eqref{svht}
\SFSnew{(where we drop the index $m$ for the sake of keeping notation simple}):
\begin{eqnarray} \label{eq:volpriors}
	\mu  \sim \mathcal{N}( b_\mu, B_\mu ), \quad \dfrac{\phi +1 }{2} \sim \mathcal{B}(a_\phi, b_\phi), \quad  \sigma^2_\eta \sim \mathcal{G}(1/2, 1/2 B_\sigma ),
\end{eqnarray}
with hyperparameters $b_\mu, B_\mu, a_\phi, b_\phi,$ and $B_\sigma$.


\section{Details on the MCMC scheme}  \label{sec_details}

\subsection{Full conditional Gibbs sampling for all variables} \label{FCG}

   Full conditional Gibbs sampling operates in the original state space representation of the model depicted in Figure~\ref{tg_graph}. 
   This MCMC sampler loops over the coefficients for $j=1, \ldots, d$ to sample the latent variables $\lambdac_{j0}, \{\psis_{jt},\lambdac_{jt},\kappaPG_{jt}\}_{t=1}^T  | \{\betaci{jt}\}_{t=0}^T, \apsi_j, \cpsi_j, \rhotr _j,\kfQc_j $ as well as the coefficient $ \rhotr _j | \{\psis_{jt},\lambdac_{jt}$, $\kappaPG_{jt}\}_{t=1}^T, \lambdac_{j0}$, $\{\betaci{jt}\}_{t=0}^T, \apsi_j, \cpsi_j$:
  \begin{enumerate}[(a)]
        
\item  For $1 \leq t \leq T$, 
sample $\kappaPG_{jt}| \lambdac_{jt}, \lambdac_{j,t-1}, \apsi_j, \cpsi_j, \rhotr _j$ from the discrete distribution given in (\ref{kappaden}).

\item Define $\kfwc^2_{jt} = \left(\betaci{jt} -  \betaci{j,t-1}\right)^2$ and, using (\ref{psi_cond_post}) for $1 \leq t \leq T$, sample $$ \psis_{jt}|\kfwc_{jt}, \lambdac_{jt}, \kfQc_j, \cpsi_j\sim \Gammainv{\cpsi_j + \frac 1 2, \frac{\kfwc^2_{jt}}{2\kfQc_j} + \lambdac_{jt}}.$$ 
        
\item For $1 \leq t \leq T-1$, sample $$ \lambdac_{jt} | \psis_{jt} ,  \kappaPG_{jt},  \kappaPG_{j,t+1}, \apsi_j, \cpsi_j, \rhotr _j \sim \Gammad{\apsi_j + \cpsi_j + \kappaPG_{jt} + \kappaPG_{j,t+1},
\frac{\apsi_j}{\cpsi_j}\frac{1 + \rhotr_j}{1- \rhotr_j } + \frac{1}{\psis_{jt} }} ,$$ see (\ref{renhgpost}); sample 
 $\lambdac_{j0} |\kappaPG_{j1}, \apsi_j, \cpsi_j, \rhotr _j$ and $\lambdac_{jT} | \psis_{jT} ,\kappaPG_{jT}, \apsi_j, \cpsi_j, \rhotr _j$ as in (\ref{renhgp0}) and 
  (\ref{renhgpT}), respectively.
  
    \item Sample $ \rhotr _j | \{\lambdac_{jt},\kappaPG_{jt}\}_{t=1}^T, \lambdac_{j0}, \apsi_j, \cpsi_j$ using an MH-step.
  \end{enumerate}
  Modifications are needed for sampling $\lambda_{jt}$ when $ t=0$ and $t=T$, namely:
\begin{eqnarray} \label{renhgp0}
  \lambdac_{j0} |\kappaPG_{j1}, \apsi_j, \cpsi_j, \rhotr _j
  \sim \Gammad{\apsi_j + \kappaPG_{j1},
  \frac{\apsi_j}{\cpsi_j}\frac{1}{1- \rhotr_j }},
  \end{eqnarray}
  and  
  \begin{eqnarray} \label{renhgpT}
    \lambdac_{jT} | \psis_{jT} ,\kappaPG_{jT}, \apsi_j, \cpsi_j, \rhotr _j
    \sim \Gammad{\apsi_j + \cpsi_j + \kappaPG_{jT},
    \frac{\apsi_j}{\cpsi_j}\frac{1}{1- \rhotr_j} + \frac{1}{\psis_{jT}}}.
    \end{eqnarray}
  Note that the sampling order is arbitrary in this sampler.


\subsection{Marginalized Gibbs sampler}

To simplify notation in Algorithm~\ref{Algo1}, let $\bm y = \left(y_1, \dots, y_T\right)$ denote the observed time series, $\bm X = (\Xbeta_1, \dots, \Xbeta_T)$ represent the observed regressors, $\bm \psis_t = \left(\psis_{1t}, \dots, \psis_{dt}\right)$ be a vector containing all local shrinkage process parameters $\psis_{jt}$ at time $t$, and $\zm$ be the set of all  latent variables and unknown model parameters in a TVP model with time-varying innovation variances. This allows us to define $\zcond{x}$ as the set of all unknowns  aside from $x$.

\newpage

	\hrule\hrule{~}
	\vspace{-13pt}
	\begin{alg}{{ \emph{MCMC inference for TVP models under the dynamic triple gamma prior.}}}\label{Algo1}
	\hrule\vspace{6pt}
		Choose starting values for all parameters except $\betai{0}, \dots, \betai{T}$  and repeat the following steps:
		\begin{enumerate} [leftmargin=2.3em, labelsep=4mm]
			\item[(a)]  Sample $\betai{0}, \dots, \betai{T}, \betaci{1}, \dots, \betaci{d}, \kfQc_1, \dots, \kfQc_d | \bm\psis_1, \dots, \bm\psis_T, \sigmaerr_1, \dots, \sigmaerr_T, \bm y, \bm X$ and all associated hyperparameters using Algorithm 1, steps (a) - (d) in \cite{cad-etal:tri}.
			\item[(b)] 
			\begin{enumerate}
			    \item[(b.1)] In the homoscedastic case 
                \SFSnew{where $\sigma_t^2=\sigmaerr$}, sample $\sigmaerr | \zcond{\sigmaerr}, \bm y, \bm X$  using Algorithm 1, step (f) in \cite{bit-fru:ach}.
			    \item[(b.2)]  
                \SFSnew{For the specification where $\sigma_t^2$ follows an SV model},
                sample $\sigmaerr_1, \dots, \sigmaerr_T,  \sigma^2_\eta, \mu, \phi | \zcond{\bm\sigma^2}, $  $ \bm y, \bm X $ as in \cite{kas-fru:anc}, for example via the R-package \texttt{stochvol} \citep{hos-kas:mod}.
			\end{enumerate}
		 \item[(c)]  For $j = 1, \dots, d$, perform the following sampling steps:
		\begin{enumerate}
		    \item[(1)] Use the approximate 
            \SFS{transition density} 
            $p(\psis_{jt} | \psis_{j,t-1}, \apsi_j, \cpsi_j, \rhotr_j)$ marginalized w.r.t. $\lambdac_{j0}, \dots, \lambdac_{j,t-1}$ and $\kappaPG_{j1},\dots, \kappaPG_{j,t-1}$ to construct the approximate conditional likelihood
		    \begin{equation*}
		        p(\psis_{j1}, \dots, \psis_{jT}|\apsi_j, \cpsi_j, \rhotr_j) \approx \prod_{t = 1}^Tp(\psis_{jt} | \psis_{j,t-1}, \apsi_j, \cpsi_j, \rhotr_j) = f(\psis_{j1}, \dots, \psis_{jT}|\apsi_j, \cpsi_j, \rhotr_j),
		    \end{equation*}to sample $\rhotr_j$ via a random walk MH step on $z = \log(\rhotr_j/(b_\rho - \rhotr_j)).$ Propose $\rhotr_j^* = b_\rho e^{z^*}/\left(1 + e^{z^*}\right)$, where $z^*\sim\Normal{z^{(m-1)}, v^2}$ and $z^{(m-1)} = \log(\rhotr_j^{(m-1)}/(b_\rho - \rhotr_j^{(m-1)}))$ depends on the previously sampled value $\rhotr_j^{(m-1)}$ of $\rhotr_j$. Accept $\rhotr_j^*$ with probability
			\begin{align*}
			\min \left\{1, \frac{ q_\rho(\rhotr_j^*)}{  q_\rho(\rhotr_j^{(m-1)}) } \right\}, \quad
		    {q_\rhotr (\rhotr_j) = f(\psis_{j1}, \dots, \psis_{jT}|\apsi_j, \cpsi_j, \rhotr_j) p(\rhotr_j)}\rhotr_j(b_\rho - \rhotr_j),
			\end{align*}
            where $p(\rhotr_j)$ is the prior \SFSnew{defined in \eqref{rho_prior}}.
			\item[(2)] For $t = 2, \dots, T - 1$, sample $\kappaPG_{jt}$ from the following discrete distribution marginalized w.r.t. $\lambdac_{j0}, \dots, \lambdac_{jT}$:
		    \begin{align*}
		       & p( \kappaPG_{jt}  | \kappaPG_{j,t-1}, \kappaPG_{j,t+1}, \psis_{jt}, \psis_{j,t-1},\rhotr_j, \apsi_j, \cpsi_j ) \\
 &\propto  \frac{\Gamfun{\apsi_j + \cpsi_j + \kappaPG_{j,t-1} + \kappaPG_{jt}}\Gamfun{\apsi_j + \cpsi_j + \kappaPG_{jt} + \kappaPG_{j,t+1}}}
{ \kappaPG_{jt} ! \Gamfun{\apsi_j  + \kappaPG_{jt}}}
\times \\
  & 
     \left[(1-\piNB _{j,t-1})\piNB_{jt} \frac{\apsi_j \psis_{jt}}{\apsi_j \psis_{jt}  + \cpsi_j (1-\rhotr _j) }\right] ^ {\kappaPG_{jt}}.
		    \end{align*}
		    Further, sample $\kappaPG_{j1}$ from 
		    \begin{align*}
  &  p( \kappaPG_{j1}  | \kappaPG_{j2}, \psis_{j1},\rhotr_j, \apsi_j, \cpsi_j ) \\
  & \propto  \frac{\Gamfun{\apsi_j + \cpsi_j + \kappaPG_{j1} + \kappaPG_{j2}}\Gamfun{\apsi_j + \kappaPG_{j1} }}{\Gamfun{\apsi_j + \kappaPG_{j1}  }
  \kappaPG_{j1} !} 
 \left[\rhotr_j \piNB_{j1}
    \frac{\apsi_j \psis_{j1}}{\apsi_j \psis_{j1}  + \cpsi_j (1-\rhotr _j) }\right] ^ {\kappaPG_{j1}},
		    \end{align*}
		    and $\kappaPG_{jT}$ from 
		    \begin{align*}
		     &p( \kappaPG_{jT}  | \kappaPG_{j,T-1},  \psis_{j,T-1}, \psis_{jT}, \rhotr_j, \apsi_j, \cpsi_j )\\
		        &\propto
\frac{\Gamfun{\apsi_j + \cpsi_j + \kappaPG_{j,T-1} + \kappaPG_{jT}}
\Gamfun{\apsi_j + \cpsi_j+ \kappaPG_{jT} }}{ \Gamfun{\apsi_j + \kappaPG_{jT} }\kappaPG_{jT} !}
    \left[ (1- \piNB_{j,T-1}) \frac{\apsi_j \psis_{jT}}{\apsi_j \psis_{jT}  + \cpsi_j (1-\rhotr _j) } \right] ^ {\kappaPG_{jT}}.
		    \end{align*}
      Efficient sampling from these densities is discussed in \SFSnew{Appendix}~\ref{sears}.
		    \item [(3)] For $t = 1, \dots, T-1$, sample $\lambdac_{jt}$ from a gamma distribution
		        \begin{align*}
    \lambdac_{jt} | \psis_{jt} ,  \kappaPG_{jt},  \kappaPG_{j,t+1}, \apsi_j, \cpsi_j, \rhotr _j
    \sim \Gammad{\apsi_j + \cpsi_j + \kappaPG_{jt} + \kappaPG_{j,t+1},
    \frac{\apsi_j}{\cpsi_j}\frac{1 + \rhotr_j}{1- \rhotr_j } + \frac{1}{\psis_{jt} }}.
    \end{align*}
    Further, for $t=0$, sample $\lambdac_{j0}$ from 
    \begin{align*}
          \lambdac_{j0} |\kappaPG_{j1}, \apsi_j, \cpsi_j, \rhotr _j
  \sim \Gammad{\apsi_j + \kappaPG_{j1},
  \frac{\apsi_j}{\cpsi_j}\frac{1}{1- \rhotr_j }},
    \end{align*}
    and $\lambdac_{jT}$ from 
    \begin{align*}
            \lambdac_{jT} | \psis_{jT} ,\kappaPG_{jT}, \apsi_j, \cpsi_j, \rhotr _j
    \sim \Gammad{\apsi_j + \cpsi_j + \kappaPG_{jT},
    \frac{\apsi_j}{\cpsi_j}\frac{1}{1- \rhotr_j} + \frac{1}{\psis_{jT}}}.
    \end{align*}
    \item[(4)] For $t = 1, \dots T$, sample $\psis_{jt}$ from an inverse gamma distribution
    \begin{align*}
          \psis_{jt} | \kfwc_{jt}, \lambdac_{jt} ,  \cpsi_j  \sim  \Gammainv{\cpsi_j + \frac{1}{2}, \lambdac_{jt} + \frac{ \kfwc_{jt}^2}{2\kfQc_j}}.
    \end{align*}
		\end{enumerate}
		\end{enumerate}
		\vspace{-6pt}

		\hrule\hrule{~}
	\end{alg}

\subsubsection{Derivation of conditional densities used in Algorithm~\ref{Algo1}}

In step (c)(1) of Algorithm~\ref{Algo1}, 
\begin{align*}
    q_\rho(\rhotr_j) &= f(\psis_{j1}, \dots, \psis_{jT}|\apsi_j, \cpsi_j, \rhotr_j)
    p(\rhotr_j)
    \rhotr_j(b_\rho - \rhotr_j) \\
    &\propto %
    \prod_{t=1}^T \biggl[  {}_2 F_1\left(\apsi_j + \cpsi_j,  \apsi_j + \cpsi_j, \apsi_j, \frac{\apsi_j^2\rhotr_j\psis_{jt}\psis_{j,t-1}}{(\apsi_j\psis_{jt} + \cpsi_j(1 - \rhotr_j))(\apsi_j\psis_{jt-1} + \cpsi_j(1 - \rhotr_j))}\right) \\
 &\times \left(\frac{(1 - \rhotr_j)}{\left(\cpsi_j + \frac{\apsi_j \psis_{jt}}{1 - \rhotr_j}\right)(\apsi_j\psis_{j,t-1} + \cpsi_j(1 - \rhotr_j))}\right)^{\apsi_j + \cpsi_j} \biggr]  
 \\
 &\times (1 - \rhotr_j)^{-T\apsi_j} \rhotr_j^{a_\rho\alpha_\rho - 1}\left(1 - \left(\rhotr_j/b_\rhotr\right)^{a_\rhotr}\right)^{\beta_\rhotr - 1}\rhotr_j(b_\rhotr - \rhotr_j).
\end{align*}
Hence, $\log q_\rho(\rhotr_j)$ is given by
\begin{align*}
    \log q_\rho(\rhotr_j) &= \biggl[\sum_{t=1}^T  \log{}_2 F_1\left(\apsi_j + \cpsi_j,  \apsi_j + \cpsi_j, \apsi_j, \frac{\apsi_j^2\rhotr_j\psis_{jt}\psis_{j,t-1}}{(\apsi_j\psis_{jt} + \cpsi_j(1 - \rhotr_j))(\apsi_j\psis_{jt-1} + \cpsi_j(1 - \rhotr_j))}\right) \\
 &+ (\apsi_j + \cpsi_j)\left(\log(1 - \rhotr_j) - \log\left(\cpsi_j + \frac{\apsi_j \psis_{jt}}{1 - \rhotr_j}\right) - \log(\apsi_j\psis_{j,t-1} + \cpsi_j(1 - \rhotr_j))\right)\biggr] \\
 &-T\apsi_j\log(1 - \rhotr_j) + (a_\rho\alpha_\rho - 1)\log\rhotr_j + (\beta_\rhotr - 1) \log\left(1 - \left(\rhotr_j/b_\rhotr\right)^{a_\rhotr}\right)\\
 &+ \log\rhotr_j + \log(b_\rhotr - \rhotr_j).
\end{align*}

For step (c)(2), it is important to note that the approximate conditional likelihood in (c)(1) is marginalized w.r.t. $\lambdac_{j0}, \dots, \lambdac_{jT}$ and $\kappaPG_{j1},\dots, \kappaPG_{jT}$. Therefore, to preserve the stationary distribution of the Markov chain, one needs to sample $\kappaPG_{j1}, \dots, \kappaPG_{jT}$ from a density that is marginalized w.r.t. $\lambdac_{j0}, \dots, \lambdac_{jT}$. As such, one has to turn to the alternate state space representation in defined by \eqref{poshflde} together with \eqref{KjtpostD} and \eqref{transKj0} and visualized in Figure~\ref{tg_graph_alt}. Hence, the posterior $\kappaPG_{jt}  | \kappaPG_{j,t-1}, \kappaPG_{j,t+1}, \psis_{jt}, \psis_{j,t-1}, \rhotr_j, \apsi_j, \cpsi_j$ is given by:
 \begin{eqnarray*} 
&&
  p( \kappaPG_{jt}  | \kappaPG_{j,t-1}, \kappaPG_{j,t+1}, \psis_{jt}, \psis_{j,t-1},\rhotr_j, \apsi_j, \cpsi_j ) \propto \\
&&  \qquad \qquad \qquad   p(\kappaPG_{jt}  | \kappaPG_{j,t-1},\psis_{j,t-1}, \apsi_j, \cpsi_j, \rhotr_j) p(\kappaPG_{j,t+1}  |  \kappaPG_{jt} , \psis_{jt}, \apsi_j, \cpsi_j, \rhotr_j) p(\psis_{jt}| \kappaPG_{jt},  \apsi_j, \cpsi_j, \rhotr_j  ).
    \end{eqnarray*}
Both transition densities 
$p(\kappaPG_{jt}  | \kappaPG_{j,t-1},\psis_{j,t-1}, \apsi_j, \cpsi_j, \rhotr_j)$
and  $ p(\kappaPG_{j,t+1}  |  \kappaPG_{jt} , \psis_{jt}, \apsi_j, \cpsi_j, \rhotr_j)$
are equal to the negative binomial transition density derived in (\ref{KjtpostD}), i.e.
\begin{eqnarray*} 
&& p( \kappaPG_{jt}  | \kappaPG_{j,t-1},  \psis_{j,t-1} , \rhotr_j, \apsi_j, \cpsi_j ) =
   \frac{\Gamfun{\apsi_j + \cpsi_j + \kappaPG_{j,t-1} + \kappaPG_{jt}}}{ \kappaPG_{jt} !  \Gamfun{\apsi_j + \cpsi_j + \kappaPG_{j,t-1}}}
   (\piNB_{j,t-1}) ^ {\apsi_j + \cpsi_j + \kappaPG_{j,t-1}} (1-\piNB _{j,t-1})^{\kappaPG_{jt}},\\
 && p( \kappaPG_{j,t+1}  | \kappaPG_{jt},  \psis_{jt} , \rhotr_j, \apsi_j, \cpsi_j ) =
   \frac{\Gamfun{\apsi_j + \cpsi_j +  \kappaPG_{jt} + \kappaPG_{j,t+1} }}{ \kappaPG_{j,t+1} !  \Gamfun{\apsi_j + \cpsi_j + \kappaPG_{jt}}}
   (\piNB_{jt}) ^ {\apsi_j + \cpsi_j + \kappaPG_{jt}} (1-\piNB_{jt})^{\kappaPG_{j,t+1}},
\end{eqnarray*}
while $p(\psis_{jt} |  \kappaPG_{jt},\apsi_j, \cpsi_j, \rhotr_j) $
   arises from the \SFS{generalized} beta prime
   distribution    derived in   (\ref{poshflde}):
    \begin{eqnarray*} 
  p( \psis_{jt} | \kappaPG_{jt}, \apsi_j, \cpsi_j, \rhotr _j ) =  
  && \frac{1}{\Betafun{\apsi_j+  \kappaPG_{jt}, \cpsi_j}}  \frac{\apsi_j}{\cpsi_j (1- \rhotr_j )} \\
  &&
\times \left( \frac{\apsi_j\psis_{jt}}{\cpsi_j (1- \rhotr_j )}  \right) ^{ \apsi_j+  \kappaPG_{jt}-1}
\left( \frac{\apsi_j\psis_{jt}}{\cpsi_j (1- \rhotr_j )} +1 \right) ^{ -({\apsi_j + \cpsi_j+  \kappaPG_{jt}})}.
\end{eqnarray*}
This yields:
\begin{eqnarray} 
   \nonumber
&&  p( \kappaPG_{jt}  | \kappaPG_{j,t-1}, \kappaPG_{j,t+1}, \psis_{jt}, \psis_{j,t-1},\rhotr_j, \apsi_j, \cpsi_j )  \\
& \propto & \frac{\Gamfun{\apsi_j + \cpsi_j + \kappaPG_{j,t-1} + \kappaPG_{jt}}}{ \kappaPG_{jt} !  \Gamfun{\apsi_j + \cpsi_j + \kappaPG_{j,t-1}}}
   (\piNB_{j,t-1}) ^ {\apsi_j + \cpsi_j + \kappaPG_{j,t-1}} (1-\piNB _{j,t-1})^{\kappaPG_{jt}} \nonumber \\
&&   \times  \frac{\Gamfun{\apsi_j + \cpsi_j + \kappaPG_{jt} + \kappaPG_{j,t+1}}}{ \kappaPG_{j,t+1} !  \Gamfun{\apsi_j + \cpsi_j + \kappaPG_{jt}}}
   (\piNB_{jt}) ^ {\apsi_j + \cpsi_j + \kappaPG_{jt}} (1-\piNB _{jt})^{\kappaPG_{j,t+1}}  \nonumber \\
&&  \times \frac{1}{\Betafun{\apsi_j+  \kappaPG_{jt}, \cpsi_j}}  \frac{\apsi_j}{\cpsi_j (1- \rhotr_j )}
\left( \frac{\apsi_j\psis_{jt}}{\cpsi_j (1- \rhotr_j )}  \right) ^{ \apsi_j+  \kappaPG_{jt}-1}
\left( \frac{\apsi_j\psis_{jt}}{\cpsi_j (1- \rhotr_j )} +1 \right) ^{ -({\apsi_j + \cpsi_j+  \kappaPG_{jt}})} \nonumber \\
& \propto & \frac{\Gamfun{\apsi_j + \cpsi_j + \kappaPG_{j,t-1} + \kappaPG_{jt}}\Gamfun{\apsi_j + \cpsi_j + \kappaPG_{jt} + \kappaPG_{j,t+1}}}
{ \kappaPG_{jt} ! \Gamfun{\apsi_j  + \kappaPG_{jt}}}
     \left[ \frac{ (1-\piNB _{j,t-1})\piNB_{jt} \apsi_j \psis_{jt}}{\apsi_j \psis_{jt}  + \cpsi_j (1-\rhotr _j) }\right] ^ {\kappaPG_{jt}}. \nonumber    \end{eqnarray}
     Modifications are needed  for $ t=1$ and $t=T$. For $ t=1$, 
    the prior  $p(\kappaPG_{j1}  |   \apsi_j, \cpsi_j, \rhotr_j)$
    is equal to  the stationary distribution    
    $\kappaPG_{j1} | \apsi_j, \rhotr_j \sim \NegBin{\apsi_j , 1-\rhotr_j }$
 given in (\ref{transKj0}) and is combined with the likelihoods $p(\kappaPG_{j2}  | \kappaPG_{j1},\psis_{j1}, \apsi_j, \cpsi_j, \rhotr_j)$ and $ p(\psis_{jT} | \kappaPG_{jT}, \apsi_j, \cpsi_j, \rhotr_j)$:
      \begin{eqnarray} \nonumber
  &&  p( \kappaPG_{j1}  | \kappaPG_{j2}, \psis_{j1},\rhotr_j, \apsi_j, \cpsi_j ) \propto   p(\kappaPG_{j2}  | \kappaPG_{j1},\psis_{j1}, \apsi_j, \cpsi_j, \rhotr_j) p(\kappaPG_{j1}  |   \apsi_j, \rhotr_j)\\
  && \qquad \propto  \frac{\Gamfun{\apsi_j + \cpsi_j + \kappaPG_{j1} + \kappaPG_{j2}}\Gamfun{\apsi_j + \kappaPG_{j1} }}{ 
  \Gamfun{\apsi_j + \kappaPG_{j1}  }
  \kappaPG_{j1} !}
    \left[\rhotr_j \piNB_{j1}
    \frac{\apsi_j \psis_{j1}}{\apsi_j \psis_{j1}  + \cpsi_j (1-\rhotr _j) }\right] ^ {\kappaPG_{j1}}. 
    \nonumber 
    \end{eqnarray}    
   For $t=T$, the \SFSnew{the negative binomial transition density} 
   $p(\kappaPG_{jT}  | \kappaPG_{j,T-1},\psis_{j,T-1}, \apsi_j, \cpsi_j, \rhotr_j)$ is combined with the likelihood 
   $p(\psis_{jT} |  \kappaPG_{jT},\apsi_j, \cpsi_j, \rhotr_j) $:
   \begin{eqnarray} 
  && p( \kappaPG_{jT}  | \kappaPG_{j,T-1},  \psis_{j,T-1}, \psis_{jT}, \rhotr_j, \apsi_j, \cpsi_j ) \propto p( \kappaPG_{jT}  | \kappaPG_{j,T-1},  \psis_{j,T-1}, 
\rhotr_j, \apsi_j, \cpsi_j ) 
p(\psis_{jT} | \kappaPG_{jT}, \apsi_j, \cpsi_j, \rhotr _j) \nonumber
\\
&& \qquad \propto
\frac{\Gamfun{\apsi_j + \cpsi_j + \kappaPG_{j,T-1} + \kappaPG_{jT}}
\Gamfun{\apsi_j + \cpsi_j+ \kappaPG_{jT} }
}{ 
\Gamfun{\apsi_j + \kappaPG_{jT} }
 \kappaPG_{jT} !}
    \left[ (1- \piNB_{j,T-1}) \frac{\apsi_j \psis_{jT}}{\apsi_j \psis_{jT}  + \cpsi_j (1-\rhotr _j) }\right] ^ {\kappaPG_{jT}}.
    \nonumber 
   \end{eqnarray} 
 Hence, the posterior $p( \kappaPG_{jt}  | 
 \kappaPG_{j,s}, \psis_{j,r}, 
\rhotr_j, \apsi_j, \cpsi_j )$, where $s \in \{ \max(t-1,1), \min(t+1,T) \}$ and  $r \in \{ \max(t-1,1), t \}$ is proportional to 
\begin{eqnarray} \label{kapnew}
p( \kappaPG_{jt}  | 
 \kappaPG_{j,s}, \psis_{j,r}, 
\rhotr_j, \apsi_j, \cpsi_j )  \propto \frac{\Gamfun{a^\kappa_{jt}+ \kappa_{jt}} \Gamfun{b^\kappa_{jt}+\kappa_{jt}}}{\Gamfun{c^\kappa_{j}+\kappa_{jt}}} \frac{(z^\psi_{jt}) ^{\kappa_{jt}}}{\kappa_{jt}!},
\end{eqnarray} 
 for $t=1, \ldots, T$, where $a^\kappa_{jt}, b^\kappa_{jt}, c^\kappa_{j}$ and $z^\psi_{jt}$ are defined in Table~\ref{tabHF}.
   It follows immediately from  the definition of the hypergeometric function presented in \eqref{hypergeo_def} that the normalizing constant of this density can be expressed with the help of the hypergeometric function, for all $t=1, \ldots, T$:
\begin{eqnarray} \label{normkap}
 \sum_{\kappa_{jt}=0}^\infty \frac{\Gamfun{a^\kappa_{jt}+ \kappa_{jt}} \Gamfun{b^\kappa_{jt}+\kappa_{jt}}}{\Gamfun{c^\kappa_{j}+\kappa_{jt}}} \frac{(z^\psi_{jt}) ^{\kappa_{jt}}}{\kappa_{jt}!}
 = \frac{\Gamfun{a^\kappa_{jt}} \Gamfun{b^\kappa_{jt}}}{\Gamfun{c^\kappa_{j}}} {}_2 F_1(a^\kappa_{jt}, b^\kappa_{jt}; c^\kappa_{j}; z^\psi_{jt}).
\end{eqnarray} 
   This allows for very efficient sampling from these densities, see \SFSnew{Appendix}~\ref{sears}.
\begin{table}
    \centering
\begin{tabular}{cccc}
    \hline
    & $t=1$ & $t=2, \ldots, T-1$ & $t=T$ \\
 \hline
     $a^\kappa_{jt}$    & $\apsi_j$ & $\apsi_j + \cpsi_j + \kappaPG_{j,t-1} $ & $\apsi_j + \cpsi_j + \kappaPG_{j,T-1}$ \\
     $b^\kappa_{jt}$    & $\apsi_j + \cpsi_j + \kappaPG_{j2}$  & $\apsi_j + \cpsi_j + \kappaPG_{j,t+1}$ & $\apsi_j + \cpsi_j$ \\ 
     $ c^\kappa_{j}$  & $\apsi_j $ & $\apsi_j $ & $\apsi_j $\\
     $z^\psi_{jt}$  & $\displaystyle \frac{\rhotr_j \piNB_{j1}\apsi_j \psis_{j1}}{\apsi_j \psis_{j1}  + \cpsi_j (1-\rhotr _j) }$
     &    $\displaystyle \frac{(1-\piNB _{j,t-1})\piNB_{jt}\apsi_j \psis_{jt}}{\apsi_j \psis_{jt}  + \cpsi_j (1-\rhotr _j) }$
     & 
     $\displaystyle \frac{(1- \piNB_{j,T-1})\apsi_j \psis_{jT}}{\apsi_j \psis_{jT}  + \cpsi_j (1-\rhotr _j) }$ \\ \hline
    \end{tabular}
    \caption{Constants required for the derivation of the function ${}_2 F_1(a^\kappa_{jt}, b^\kappa_{jt}; c^\kappa_{j}; z^\psi_{jt}) $ for the normalising constant of $ \kappaPG_{jt}  | \kappaPG_{j,t-1}, \kappaPG_{j,t+1}, \psis_{jt}, \psis_{j,t-1},\rhotr_j, \apsi_j, \cpsi_j$. Note that $\piNB_{jt}= \frac{\apsi_j\psis_{jt} + \cpsi_j (1- \rhotr_j) }{(1+\rhotr_j) \apsi_j\psis_{jt} + \cpsi_j (1- \rhotr_j) }$, as in \eqref{KjtpostD}. }\label{tabHF}
\end{table}

Step (c)(3) is more straightforward, as we can simply sample from the full conditional distribution again.
For $t = 1, \dots, T-1$:
\begin{align*}
    p(\lambdac_{jt}| \psis_{jt},  \kappaPG_{jt},  \kappaPG_{j,t+1}, \apsi_j, \cpsi_j, \rhotr _j) &\propto p(\psis_{jt}|\lambdac_{jt}, \apsi_j, \cpsi_j, \rhotr_j)p(\kappaPG_{j,t+1}|\lambdac_{jt}, \apsi_j, \cpsi_j, \rhotr_j)p(\lambdac_{jt}|\kappaPG_{jt}, \apsi_j, \cpsi_j, \rhotr_j) \\
    &\propto \lambdac_{jt}^{\cpsi_j}e^{-\frac{\lambdac_{jt}}{\psis_{jt}}} \lambdac_{jt}^{\kappaPG_{j,t+1}}e^{-\frac{\apsi_j}{\cpsi_j}\frac{\rhotr_j}{1 - \rhotr_j}\lambdac_{jt}} \lambdac_{jt}^{\apsi_j + \kappaPG_{jt} - 1}e^{-\frac{\apsi_j}{\cpsi_j}\frac{1}{1 - \rhotr_j}\lambdac_{jt}} \\
    &= \lambdac_{jt}^{\apsi_j + \cpsi_j + \kappaPG_{jt} + \kappaPG_{j,t+1} - 1}e^{-\lambdac_{jt}\left(\frac{\apsi_j}{\cpsi_j}\frac{1 + \rhotr_j}{1 - \rhotr_j} + \frac{1}{\psis_{jt}}\right)},
\end{align*}
which is the kernel of a $\Gammad{\apsi_j + \cpsi_j + \kappaPG_{jt} + \kappaPG_{j,t+1},
\frac{\apsi_j}{\cpsi_j}\frac{1 + \rhotr_j}{1- \rhotr_j } + \frac{1}{\psis_{jt} }}$-distribution.
Modifications are needed  for $ t=0$ and $t=T$. For $t = 0$, note that:
\begin{align*}
    p(\lambdac_{j0}| \kappaPG_{j1}, \apsi_j, \cpsi_j, \rhotr _j) &\propto p(\kappaPG_{j1}|\lambdac_{j0}, \apsi_j, \cpsi_j, \rhotr_j)p(\lambdac_{j0}| \apsi_j, \cpsi_j) \\
    &\propto \lambdac_{j0}^{\kappaPG_{j1}}e^{-\frac{\apsi_j}{\cpsi_j}\frac{\rhotr_j}{1 - \rhotr_j}\lambdac_{j0}} \lambdac_{j0}^{\apsi_j - 1}e^{-\frac{\apsi_j}{\cpsi_j}\lambdac_{j0}} \\
    &= \lambdac_{j0}^{\apsi_j + \kappaPG_{j1} - 1}e^{-\frac{\apsi_j}{\cpsi_j}\frac{1}{1 - \rhotr_j}\lambdac_{j0}} ,
\end{align*}
which is the kernel of a $\Gammad{\apsi_j + \kappaPG_{j1},
  \frac{\apsi_j}{\cpsi_j}\frac{1}{1- \rhotr_j }}$-distribution. For $t = T$, note that
\begin{align*}
    p(\lambdac_{jT}| \psis_{jT}, \kappaPG_{jT}, \apsi_j, \cpsi_j, \rhotr _j) &\propto p(\psis_{jT}|\lambdac_{jT}, \cpsi_j)p(\lambdac_{jT}| \kappaPG_{jT}, \apsi_j, \cpsi_j, \rhotr_j) \\
    &\propto \lambdac_{jT}^{\cpsi_j}e^{-\frac{\lambdac_{jT}}{\psis_{jT}}} \lambdac_{jT}^{\apsi_j + \kappaPG_{jT} - 1}e^{-\frac{\apsi_j}{\cpsi_j}\frac{1}{1 - \rhotr_j}\lambdac_{jT}} \\
    &= \lambdac_{jT}^{\apsi_j + \cpsi_j + \kappaPG_{jT} - 1}e^{-\lambdac_{jT}\left(\frac{\apsi_j}{\cpsi_j}\frac{1}{1 - \rhotr_j} + \frac{1}{\psis_{jT}}\right)} ,
\end{align*}
  which is the kernel of a $\lambdac_{jT} | \psis_{jT} ,\kappaPG_{jT}, \apsi_j, \cpsi_j, \rhotr _j
    \sim \Gammad{\apsi_j + \cpsi_j + \kappaPG_{jT},
    \frac{\apsi_j}{\cpsi_j}\frac{1}{1- \rhotr_j} + \frac{1}{\psis_{jT}}}$-distribution.

Finally, in step (c)(4)  $\psis_{jt}| \lambdac_{jt},\kfwc_{jt},\cpsi_j $ follows an inverse gamma distribution, as
\begin{align*}
    p(\psis_{jt} | \kfwc_{jt}, \lambdac_{jt} ,  \cpsi_j) &\propto p(\kfwc_{jt}|\kfQc_j, \psis_{jt})p(\psis_{jt}|\lambdac_{jt}, \cpsi_j) \\
    &\propto \psis_{jt}^{-\frac 1 2} e^{-\frac{\kfwc_{jt}^2}{2\kfQc_j\psis_{jt}}} \psis_{jt}^{-\cpsi_j - 1} e^{-\frac{\lambdac_{jt}}{\psis_{jt}}} \\
    &= \psis_{jt}^{-\left(\cpsi_j + \frac 1 2\right) - 1} e^{-\frac{1}{\psis_{jt}}\left(\lambdac_{jt} + \frac{\kfwc_{jt}^2}{2\kfQc_j}\right)}
\end{align*}
which is the kernel of an $\psis_{jt} | \kfwc_{jt}, \lambdac_{jt} ,  \cpsi_j  \sim  \Gammainv{\cpsi_j + \frac{1}{2}, \lambdac_{jt} + \frac{ \kfwc_{jt}^2}{2\kfQc_j}}$-distribution.

\subsubsection{Efficient sampling of $\kappaPG_{jt}$} \label{sears}

To sample $\kappaPG_{jt}$ in step~(c)(2) in Algorithm~\ref{Algo1}, we need to sample from a $\dimmat{d}{T}$-dimensional discrete random variable, where $\pkappa_{jt,k}=\Prob{\kappaPG_{jt}=k| \kappaPG_{j,s}, \psis_{j,r}, \rhotr_j, \apsi_j, \cpsi_j}$, for $s \in \{ \max(t-1,1), \min(t+1,T) \}$ and $r \in \{ \max(t-1,1), t \}$, is given by the unnormalized weights provided in \eqref{kapnew}, divided by the normalization constant given in \eqref{normkap}.

 To this end, we can exploit that $\pkappa_{jt,k}$ satisfies a recursive scheme. The initial value $\pkappa_{jt,0}$ is given by:
\begin{eqnarray} \label{reptinit}
\pkappa_{jt,0} =  \frac{1}{{}_2 F_1(a^\kappa_{jt}, b^\kappa_{jt}; c^\kappa_{j}; z^\psi_{jt}) },
\end{eqnarray}
with succesive values for $k \geq 1$ given by, 
\begin{eqnarray} \label{recurs}
    \pkappa_{jt,k} = \pkappa_{jt,k-1} \kappaint_{jt,k},
\quad \kappaint_{jt,k}=  \frac{(a^\kappa_{jt}+k-1) (b^\kappa_{jt}+k-1)}{(c^\kappa_{j} +k-1)} \frac{z^\psi_{jt}}{k}.
\end{eqnarray}
  This makes it possible to sample the entire $\dimmat{d}{T}$-dimensional discrete random variable $\kappaM=\{\{\kappaPG_{jt}\}_{j=1}^{d} \}_{t=1}^T$ with a recursive algorithm that does not require precomputation of $\pkappa_{jt,k}$ 
  \SFS{except for $k=0$}:
  \begin{enumerate}[(a)]
    \item Sample a $\dimmat{d}{T}$-array $\Um$ of uniform random numbers 
    \SFS{with elements $\Um_{jt}$}
    and
    determine the $\dimmat{d}{T}$-array $\pkappam^{(0)}$ of initial probabilities
     \SFS{with each element $ \pkappam^{(0)}_{jt} =\pkappa_{jt,0}$ defined in \eqref{reptinit}}.
    If  $\Um_{jt} \leq \pkappam^{(0)}_{jt} $, then $\kappaPG_{jt}=0$.
    \item Further calculations are only required for the elements
    \SFS{$\kappaPG_{jt}$} in $\kappaM$ where  $\Um_{jt} > \pkappam^{(0)}_{jt} $. For these elements, $\kappaPG_{jt}$ is at least equal to 1. We determine $\pkappam^{(1)}$ from $\pkappam^{(0)}$ using (\ref{recurs}).
        If $\Um_{jt} <\pkappam^{(0)}_{jt} + \pkappam^{(1)} _{jt} $, then $\kappaPG_{jt}=1$.
      \item Further calculations are only required for the elements in $\kappaM$ where  $\Um_{jt} > \pkappam^{(0)}_{jt}  + \pkappam^{(1)} _{jt}$. For these elements, $\kappaPG_{jt}$ is at least equal to 2.
         \end{enumerate}
 This leads to following recursion: starting with $\kappaM ^{(0)}=\bfzmat$, $\Um ^{(0)}=\Um$, $\pkappam^{(0)}$ and $k=0$, repeat the following steps:
         \begin{enumerate}[(a)]
        \item Increase $k$ by 1.
          \item Update  $ \kappaM ^{(k)} = \kappaM ^{(k-1)}  + \indic{\Um ^{(k-1)}  > \pkappam^{(k-1)}}$,
            where the indicator function is applied element-wise.
           \item  Update $ \Um ^{(k)} = \Um ^{(k-1)} - \pkappam^{(k-1)}$.
          \item Update  $\pkappam^{(k)}=\pkappam^{(k-1)} \odot \Phim ^{(k)}$,
             where the elements $\kappaint_{jt,k}$ of the $\dimmat{d}{T}$ matrix $\Phim ^{(k)}$ are given in 
        \eqref{recurs} and
        $\odot$ is element-wise matrix multiplication.
        \item Terminate the recursion if $\Um ^{(k)} < \pkappam^{(k)}$ for all elements or $k>\Kmax$.
       \end{enumerate}

\section{Data overview} \label{data}

Table~\ref{data_desc} gives an overview of the companies that comprise the EURO STOXX 50 index. Importantly, the tickers (e.g. FLTR.IR) used in the plots in Section~\ref{appl} are connected to the company names, as the connection between the two is often non-obvious.  The companies shaded in red were excluded from the application, either due to the number of missing values or because the data on the returns did not reach back far enough. 

\begin{table}
\centering
\resizebox{0.47\textwidth}{!}{

\begin{tabular}[t]{lll}
\toprule
Ticker & Name & Reg. office\\
\midrule
ADS.DE & Adidas & Germany\\
\cellcolor[HTML]{FAA0A0}{ADYEN.AS} & \cellcolor[HTML]{FAA0A0}{Adyen} & \cellcolor[HTML]{FAA0A0}{Netherlands}\\
\cellcolor[HTML]{FAA0A0}{AD.AS} & \cellcolor[HTML]{FAA0A0}{Ahold Delhaize} & \cellcolor[HTML]{FAA0A0}{Netherlands}\\
AI.PA & Air Liquide & France\\
AIR.PA & Airbus & France\\
\addlinespace
ALV.DE & Allianz & Germany\\
ABI.BR & Anheuser-Busch InBev & Belgium\\
ASML.AS & ASML Holding & Netherlands\\
CS.PA & AXA & France\\
BAS.DE & BASF & Germany\\
\addlinespace
BAYN.DE & Bayer & Germany\\
BBVA.MC & BBVA & Spain\\
SAN.MC & Banco Santander & Spain\\
BMW.DE & BMW & Germany\\
BNP.PA & BNP Paribas & France\\
\addlinespace
CRG.IR & CRH & Ireland\\
BN.PA & Danone & France\\
DB1.DE & Deutsche Börse & Germany\\
DPW.DE & Deutsche Post & Germany\\
DTE.DE & Deutsche Telekom & Germany\\
\addlinespace
ENEL.MI & Enel & Italy\\
ENI.MI & Eni & Italy\\
EL.PA & EssilorLuxottica & France\\
FLTR.IR & Flutter Entertainment & Ireland\\
RMS.PA & Hermès & France\\
\bottomrule
\end{tabular}
}
\hspace{0.5em}
\resizebox{0.47\textwidth}{!}{

\begin{tabular}[t]{lll}
\toprule
Ticker & Name & Reg. office\\
\midrule
IBE.MC & Iberdrola & Spain\\
ITX.MC & Inditex & Spain\\
IFX.DE & Infineon Technologies & Germany\\
INGA.AS & ING Group & Netherlands\\
ISP.MI & Intesa Sanpaolo & Italy\\
\addlinespace
KER.PA & Kering & France\\
KNEBV.HE & Kone & Finland\\
OR.PA & L'Oréal & France\\
\cellcolor[HTML]{FAA0A0}{LIN.DE} & \cellcolor[HTML]{FAA0A0}{Linde plc} & \cellcolor[HTML]{FAA0A0}{Ireland}\\
MC.PA & LVMH & France\\
\addlinespace
MBG.DE & Mercedes-Benz Group & Germany\\
MUV2.DE & Munich Re & Germany\\
RI.PA & Pernod Ricard & France\\
PHIA.AS & Philips & Netherlands\\
\cellcolor[HTML]{FAA0A0}{PRX.AS} & \cellcolor[HTML]{FAA0A0}{Prosus} & \cellcolor[HTML]{FAA0A0}{Netherlands}\\
\addlinespace
SAF.PA & Safran & France\\
SAN.PA & Sanofi & France\\
SAP.DE & SAP & Germany\\
SU.PA & Schneider Electric & France\\
SIE.DE & Siemens & Germany\\
\addlinespace
STLA.MI & Stellantis & Netherlands\\
TTE.PA & TotalEnergies & France\\
DG.PA & Vinci SA & France\\
VOW.DE & Volkswagen Group & Germany\\
\cellcolor[HTML]{FAA0A0}{VNA.DE} & \cellcolor[HTML]{FAA0A0}{Vonovia} & \cellcolor[HTML]{FAA0A0}{Germany}\\
\bottomrule
\end{tabular}
}
\caption{\label{data_desc}Tickers, names and registered offices of companies that comprise the EURO STOXX 50 index. Companies shaded red were excluded from the application in Section~\ref{appl} due to data availability issues.}
\end{table}

\end{appendices}
\end{document}